\documentclass{aastex631}

\usepackage{longtable}
\usepackage{rotating}
\usepackage{lineno}
\usepackage{epsfig}

\begin{document}
\title{The Thermal Emission in Short Gamma-Ray Bursts with Extended Emission Observed
by Fermi/GBM}

\author{Xue-Zhao Chang}
\affiliation{Guangxi Key Laboratory for
Relativistic Astrophysics, School of Physical Science and Technology, Guangxi University,
Nanning 530004, China; lhj@gxu.edu.cn}

\author{Hou-Jun L\"{u}$^\ast$}
\affiliation{Guangxi Key Laboratory for
Relativistic Astrophysics, School of Physical Science and Technology, Guangxi University,
Nanning 530004, China; lhj@gxu.edu.cn}

\author{Xing Yang}
\affiliation{Guangxi Key Laboratory for
Relativistic Astrophysics, School of Physical Science and Technology, Guangxi University,
Nanning 530004, China; lhj@gxu.edu.cn}

\author{Jia-Ming Chen}
\affiliation{School of Physics and Astronomy, Yunnan University, Kunming 650500, China}

\author{En-Wei Liang}
\affiliation{Guangxi Key Laboratory for
Relativistic Astrophysics, School of Physical Science and Technology, Guangxi University,
Nanning 530004, China; lhj@gxu.edu.cn}

\begin{abstract}
Short gamma-ray bursts (SGRBs) with extended emission (EE) are composed of initial main emission (ME) with a short-hard spike, followed by a long-lasting EE. Whether the ME and EE originated from the same origin or not, as well as the jet composition, remains an open question. In this paper, we present a systematic analysis of 36 gamma ray bursts (GRBs) in our sample, which are identified as the category of SGRBs with EE as observed by Fermi/Gamma-ray Burst Monitor. By extracting time-integrated spectra of ME and EE with cutoff power-law or Band models for our sample, we find that 20 out of 36 SGRBs have $\alpha$ values that exceed the death line (e.g., -2/3) of synchrotron emission within either ME or EE phases, and we suggest that the quasi-thermal component should exist in the prompt emission. Then, we extract the time-resolved spectra of our samples, but only four GRBs are bright enough to extract the time-resolved spectra. We find that both thermal and nonthermal emissions do exist in the prompt emission of those four bright GRBs, which suggests that a hybrid jet (e.g., matter and Poynting-flux outflow) in GRBs should exist. Moreover, strong positive correlations (e.g., $F_{\rm tot}-\Gamma$ and $F_{\rm tot}-kT$) in the time-resolved spectra of ME and EE for those four GRBs have been discovered. This indicates that the spectral evolution of both ME and EE seem to share similar behavior, possibly from the same physical origin.

\end{abstract}

\section{Introduction} \label{sec:intro}
Nearly 50 years after the announcement of the discovery of gamma-ray bursts (GRBs), great progress was made in both observations and theories \citep{2015PhR...561....1K}. Phenomenally, based on the duration, GRBs are divided into long-duration GRBs (LGRBs) and short-duration GRBs (SGRBs) with a division line at the observed duration $T_{\rm 90}=2$ s \citep{1993ApJ...413L.101K}. Robust associations of the underlying supernovae (SNe) with some LGRBs and the fact that LGRB host galaxies are in intense star formation suggest that they are likely to be from the core collapse of a massive star \citep{1998Natur.395..670G,2003ApJ...591L..17S,2006Natur.441..463F}, while the host galaxies of SGRBs, which exhibit little star formation and are associated with kilonovae, indicate that SGRBs originate from the merger of two compact stars, such as neutron star-neutron star (NS-NS) and neutron star-black hole (NS-BH) mergers (see \citealt{2014ARA&A..52...43B} for review). However, there is a small fraction of GRBs whose light curve of prompt emission is composed of a short-hard spike followed by a long-lasting, soft extended emission. Their burst duration is longer than 2s but possibly originates from compact star mergers, such as the first GRB 060614 \citep{2006Natur.444.1044G,2006Natur.444.1053G,2015NatCo...6.7323Y}. We named those GRBs as short GRBs with extended emission \citep{2015MNRAS.452..824K,2020MNRAS.492.3622L}, or long-short GRBs \citep{2015NatCo...6.7323Y}. Whether from massive star collapse or compact star mergers, a black hole or neutron star may be formed as the central engine after the catastrophic event \citep{Usov1992,Thompson1994,Dai1998a,Dai1998b,1999ApJ...518..356P,Narayan2001,Zhang2001,2006ApJ...643L..87G,2013ApJ...765..125L,Lv2014,Lv2015}. An ultrarelativistic jet is launched from the central engine, and powers the observed prompt emission of GRB via photosphere emission, internal shock, or magnetic dissipation\citep{1994ApJ...430L..93R,2000ApJ...530..292M,2005ApJ...628..847R,2009ApJ...700L..65Z,2010MNRAS.407.1033B,2010AIPC.1279...28I,2010ApJ...725.1137L,2011ApJ...732...49P,2011ApJ...726...90Z}. The broadband afterglow emission is attributed to synchrotron emission from the external shock when the fireball is decelerated by a circumburst medium \citep{1997ApJ...476..232M,1998ApJ...497L..17S}.

Traditionally, two scenarios of jet composition have been proposed to interpret the observations of GRB prompt emission. One is a matter-dominated fireball with a quasi-thermal component at the photosphere radius(photosphere model; \citealt{1986ApJ...308L..43P,1986ApJ...308L..47G,1990ApJ...365L..55S,1993ApJ...415..181M,1993MNRAS.263..861P}) and a nonthermal component produced by internal shocks at a larger radius(internal shock model; \citealt{1994ApJ...430L..93R,1997ApJ...490...92K,1998MNRAS.296..275D,2006AIPC..836..181P,2014AcASn..55..354Z}). The other one is a nonthermal component via the internal-collision-induced magnetic reconnection and turbulence (ICMART) from the Poynting-flux-dominated outflow  \citep{2011ApJ...726...90Z}. On the other hand, the radiation mechanism of GRB remains an open question \citep{2011CRPhy..12..206Z}. Several radiation mechanisms are also proposed to interpret the nonthermal emission of GRB, such as jitter radiation \citep{2000ApJ...540..704M}, synchrotron self-Compton \citep{2000ApJ...544L..17P}, as well as hadronic mechanisms \citep{2007MNRAS.380...78G,2009ApJ...705L.191A}.

From the observational point of view, the purely quasi-thermal component predicted by the fireball model and purely nonthermal emission from the magnetic dissipation were observed in GRB 090902B\citep{2009ApJ...706L.138A,2010ApJ...709L.172R,2011ApJ...730..141Z} and GRB 080916C \citep{2009ApJ...700L..65Z}, respectively. Moreover, a handful of GRBs have been discovered  to contain both thermal and nonthermal components in the prompt emission, such as GRB 081221 \citep{2018ApJ...866...13H}, GRB 100724B \citep{2011ApJ...727L..33G}, GRB 110721A \citep{2012ApJ...757L..31A}, GRB 120323A \citep{2013ApJ...770...32G}, GRB 160625B \citep{2017ApJ...849...71L,2017Natur.547..425T,2019ApJS..242...16L}, GRB 190114C \citep{2023ApJ...944L..57L} and GRB 210610B\citep{2022ApJ...932...25C}. Those observed solid cases suggest that the GRB jet composition is likely diverse but is still difficult to diagnose for most GRBs from observational data \citep{2015ApJ...801...2,2018NatAs...2...69Z,2020ApJ...894..100L}.

Recently, the prompt emission of several peculiar GRBs observed by both Swift and Fermi is found to consist of a short-hard spike followed by a long-lasting, soft extended emission (EE). So they are named SGRBs with EE \citep{2009ApJ...703.1696Z}, such as GRB 211227A \citep{2022ApJ...931L..23L}, GRB 211211A \citep{2022Natur.612..223R,2022Natur.612..232Y,2022Natur.612..228T,2023ApJ...943..146C,2023NatAs...7...67G}, and GRB 230307A \citep{2023arXiv230705689S,2023arXiv231007205Y,2024MNRAS.529L..67D,2024ApJ...962L..27D}. The feature of prompt emission of those GRBs is quite similar to that of GRB 060614, which was the first to be defined as SGRB with EE \citep{2006Natur.444.1044G,2006Natur.444.1053G, 2006Natur.444.1047F, 2006Natur.444.1050D}, and they are possibly from the merger of compact stars, supported by several lines of observational evidence. Nevertheless,  the long duration of SGRBs with extended emission is very difficult to explain by invoking NS–NS or NS-BH mergers, but it seems to be consistent with the merger of a white dwarf-neutron star (WD-NS) due to the large radius of the white dwarf \citep{2022Natur.612..232Y,2023ApJ...947L..21Z}. More interestingly, \citet{2023ApJ...943..146C} found that a potential hybrid jet includes both nonthermal and thermal emissions in GRB 211211A. However, another case, GRB 230307A, was thought to dominate by a Poynting-flux outflow \citep{2023arXiv231007205Y,2024MNRAS.529L..67D}. So, understanding the progenitor and jet composition of the cases of SGRBs with EE has attracted great attention in GRB studies.

Motivated by analysis of the jet composition of GRB 211211A, several questions emerge. What is the jet composition of SGRBs with EE in the  Fermi/Gamma-ray Burst Monitor (GBM) catalog? What is the fraction of SGRBs with EE that exists with thermal emission? To answer these questions, it is very important to understand the physical processes and mechanisms of SGRBs with EE, especially if they originate from the merger of compact stars \citep{2011ApJ...730..141Z,2015PhR...561....1K,2018pgrb.book.....Z}. In fact, previous works also searched for cases of SGRBs with EE in Swift \citep{2009ApJ...703.1696Z,2010ApJ...717..411N,2011ApJS..195....2S} and Fermi \citep{2015MNRAS.452..824K,2018ApJ...862..155L,2020MNRAS.492.3622L}. However, they only focused on the time-integrated spectra instead of the time-resolved spectra of prompt emission and ignored the problem of low-energy spectrum index ($\alpha$) of Band or cutoff power-law (CPL) functions exceeding the death line\footnote{For the slow-cooling synchrotron emission scenario, the maximum value of $\alpha$ cannot be larger than -2/3 which is referred to as the "death line of synchrotron emission" \citep{1998ApJ...506L..23P}.} of synchrotron if we believe that the nonthermal emission originates from synchrotron emission.

In this paper, we have systematically searched for the SGRBs with EE in the Fermi catalog to perform both time integrated and detailed time-resolved spectra analysis. We try to find out whether the quasi-thermal component exists in other SGRBs with EE, and how the quasi-thermal component evolves. The methodology of the data analysis is shown in Section \ref{s2}. In Section \ref{s3}, we present the results of spectral fits and evolution. The conclusions are drawn in Section \ref{s4} with some additional discussion.

\section {Data Reduction and Sample Selection}{\label{s2}}
The adopted data of this task are taken from the GBM  \citep{2009ApJ...702..791M} of the Fermi Gamma-ray Space Telescope. The GBM contains 12 sodium iodide (NaI) detectors that cover an energy range from 8 keV to 1 MeV and two bismuth germanate (BGO) scintillation detectors that are sensitive to higher energies between 200 keV and 40 MeV.

\subsection{Light-curve extraction}
The original GBM data of 12 NaI and 2 BGO detectors can be obtained from the public science support center at the official Fermi website\footnote{http://fermi.gsfc.nasa.gov/ssc/data/}. There are three types of data modes for each of the 14 GBM detectors, the continuous time (CTIME) data, the continuous spectroscopy (CSPEC) data, and the time-tagged event (TTE) data. CTIME data and CSPEC data correspond to a time resolution of 64 ms and 1.024 s, respectively. Moreover, the TTE data are the type of unbinned data that comprise individually digitalized pulse height resolution events from the GBM detector during a burst event. It provides an energy resolution of 128 channels and records the time interval of photons ranging from -20 to 300s. Any bin size of time resolution can be used to perform both spectral and temporal analysis. We select the data from all the bright NaI detectors and the brightest BGO detector to perform the analysis. The standard response files provided by the GBM team are used in this work. Then, we develop a {\em Python} code to extract the energy-dependent light curves and time-dependent spectra using the spectrum source package {\em Fermi ScienceTools}. More details can be found in our previous papers \citet{2018ApJ...862..155L,2020MNRAS.492.3622L}.

\subsection{Sample selection}
As of 2024 January , we have extracted the light curves of 3300 that were detected by Fermi/GBM. The motivation of this work is to study the jet composition of SGRBs with EE and the fraction of SGRBs with EE that exhibit the thermal emission. So, the sample selection is the first step in studying the abovementioned interesting questions. GRB 060614 was the first to be defined in the category of SGRB with EE, and its light curve of prompt emission is composed of a short spike lasting $\sim$5 s, followed by a longer soft emission \citep{2006Natur.444.1044G,2006Natur.444.1053G,2006Natur.444.1047F,2006Natur.444.1050D}. Afterwards, more solid cases are identified as SGRBs with EE \citep{2020MNRAS.492.3622L}, such as GRB 211227A \citep{2022ApJ...931L..23L} and GRB 230307A \citep{2023arXiv230705689S,2023arXiv231007205Y,2024MNRAS.529L..67D,2024ApJ...962L..27D}. Especially, the most sought-after GRB 211211A has a light curve of prompt emission composed of an initial hard-main emission lasting $\sim$13 s, followed by a series of soft emissions \citep{2022Natur.612..223R,2022Natur.612..232Y,2022Natur.612..228T,2023ApJ...943..146C,2023NatAs...7...67G}.

We adopt the properties of the light curve of GRB 211211A as the 'standard event' to search for the sample in this work. Two criteria of our sample selection can be described as follows:
(i) the duration of the initial main emission (ME) is less than 13 s and is followed by a longer soft emission lasting a few seconds to hundreds of seconds;
(ii) the signal-to-noise ratio (S/N) of the initial ME and EE components should be greater than 8 and 5, respectively. There are 36 GRBs that satisfy our criteria (see Table \ref{Tablel}). Our samples show overlap with \citet{2020MNRAS.492.3622L} but extend to 36 GRBs. The duration of ME ranges from 0.5 to 12.5 s, the duration of EE ranges from 2.9 to 63 s, and the whole duration of the bursts ranges from 4.3 to 65 s. Then, we focus on performing both time-integrated and time-resolved spectral analyses for our samples, and the details of the analysis are shown in Section \ref{s3}.

\begin{figure}[htb]
\centering
 \includegraphics [angle=0,scale=0.32] {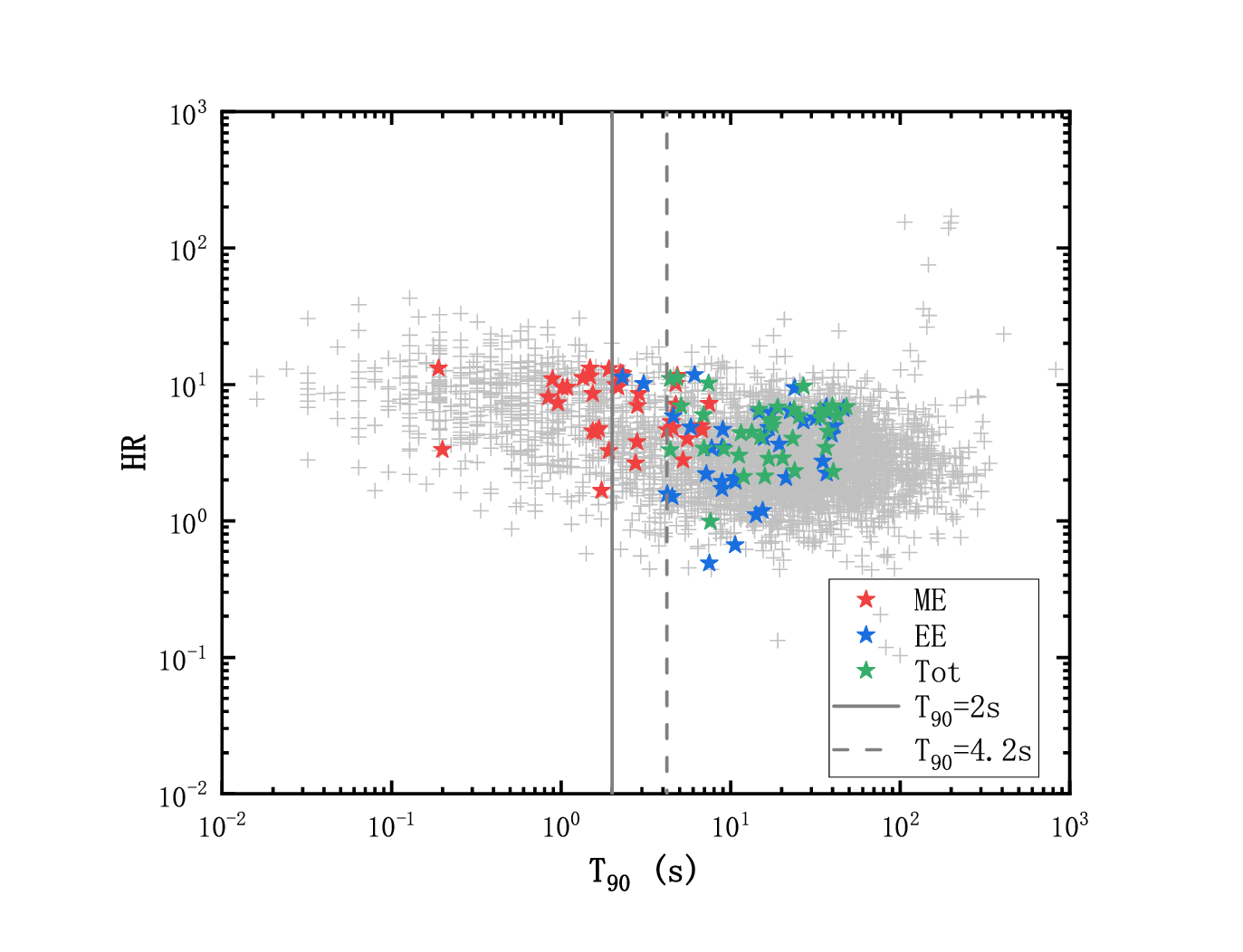}
 \includegraphics [angle=0,scale=0.35] {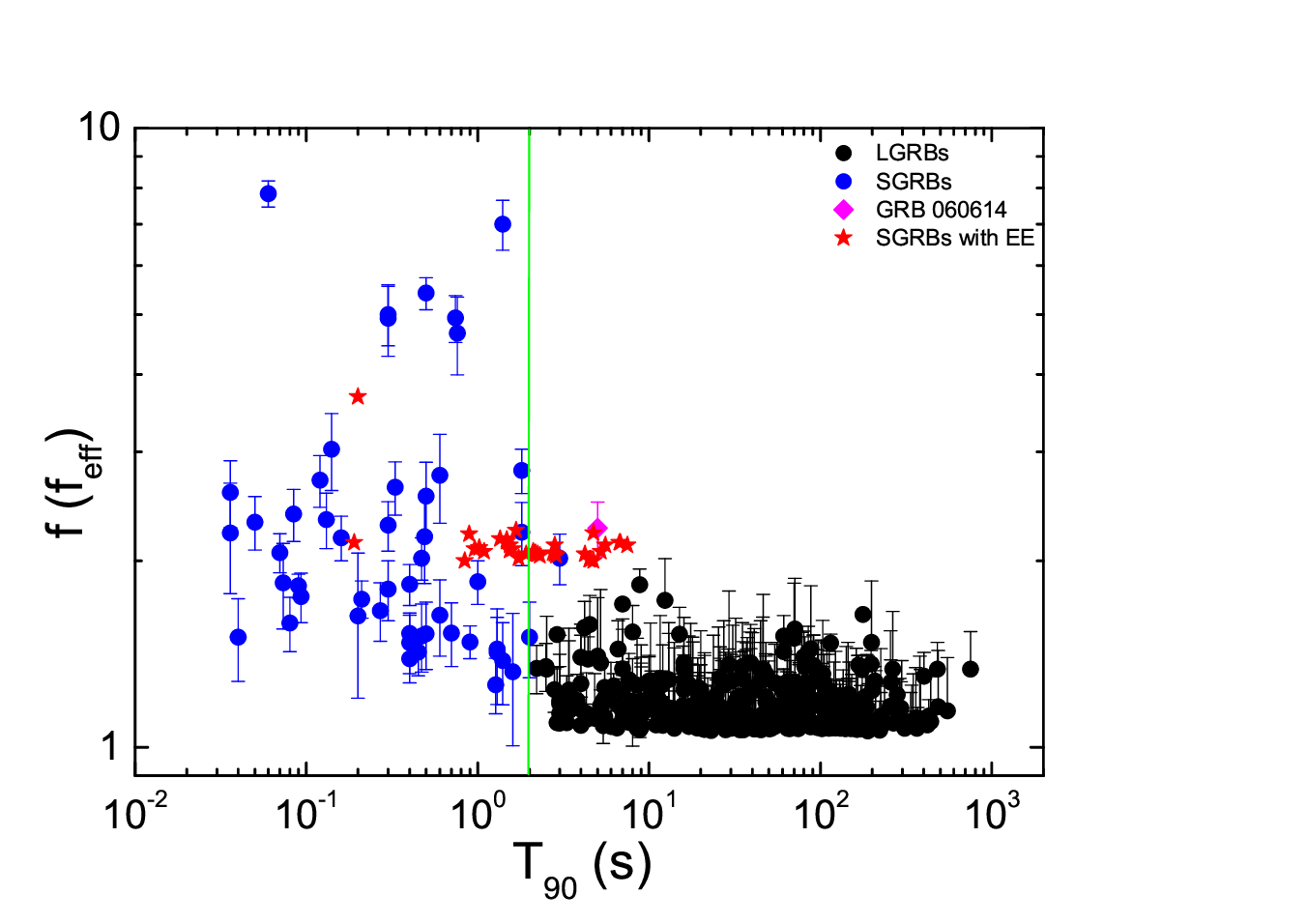}
 \caption{Left: HR vs. $T_{90}$ diagram of our sample and other long- and short-GRBs. The solid and dashed lines are the dividing lines of long and short GRBs from BATSE and Fermi \citep{2020ApJ...893...46V} catalogue, respectively. Right: The $f_{\mathrm{eff}}$ vs. $T_{90}$ diagram. The gray vertical line is the division line at 2 s. The data of long- and short-GRBs are taken from \citet{2014MNRAS.442.1922L}. GRB 060614 and our SGRBs with EE samples are highlighted by the purple square and the red star, respectively.}
 \label{fig:hr-t90}
\end{figure}

On the other hand, in order to compare the observational properties of our sample with those of other long- and short-duration GRBs, some comparisons need to be done, such as specific star-forming rate (SFR), location offset, duration, hardness, $E_{\rm p}-E_{\rm iso}$, $E_{\rm p}-L_{\rm iso}$, energetics, collimation, afterglow properties, redshift distribution, and luminosity function \citep{2009ApJ...703.1696Z}. Unfortunately, due to lack of redshift measurements for our sample, we can only plot the two-dimensional diagram of hardness ratio (HR) versus $T_{90}$ in Figure \ref{fig:hr-t90}. We find that the SGRB with EE in our sample indeed belong to the category of long-duration GRBs in terms of the total duration. However, the ME component of our sample exhibits significant differences from traditional LGRBs (seems to be closer to SGRBs in distribution), while the EE component is more akin to the distribution of LGRBs.

Moreover, \citet{2014MNRAS.442.1922L} proposed a new phenomenological classification method for GRBs by introducing a amplitude parameter $f=F_{\mathrm{p}} / F_{\mathrm{b}}$, where $F_{p}$ and $F_{b}$ represent the peak flux and background flux during the same epoch, respectively. In order to distinguish whether the short-duration GRBs are the tip of the iceberg of LGRBs, they defined an effective amplitude parameter $f_{\mathrm{eff}}=F_{\mathrm{p}}^{\prime} / F_{\mathrm{b}}$, where $F_{\mathrm{p}}^{\prime}$ is the peak flux of pseudo-GRB, which is rescaled down from an original GRB light curve until its signal above the background has a duration shorter than 2 s. Following the calculation procedure presented in \citet{2014MNRAS.442.1922L}, we obtain the effective amplitudes for our SGRBs with EE sample. Then, we plot the diagram of $f_{\mathrm{eff}}$ as function of duration in Figure \ref{fig:hr-t90}. It is found that the parameters of $f_{\mathrm{eff}}$ for our sample are significantly higher than those of other long-duration GRBs, but are consistent with those of other short-duration GRBs or SGRB with EE (such as GRB 060614). This suggests that the SGRBs with EE of our sample seem to originate from the merger of compact stars, even though the duration is longer.

\subsection{Spectral model selected and spectral fitting}
In our analysis, several spectral models can be selected to do the spectral fitting of the burst, such as (1) Band function (Band), (2) CPL, (3) blackbody (BB), (4) power-law (PL) models, as well as combinations of any two models. Those models are adopted to do the fits of spectra in GRB studies \citep{2011ApJ...730..141Z,2015ApJ...807..148G}. The details of above models are shown as follows.

(1). The Band function \citep{1993ApJ...413..281B} model can be written as,
 \begin{eqnarray}
N_{\textrm{Band}}(E)=A\left\{\begin{array}{clcc}
(\frac{E}{100~\mathrm{keV}})^{\alpha }\mathrm{exp}\left[-\frac{(\alpha+2)E}{E_{p}} \right ], E \leq (\alpha-\beta) E_{c}, \\
(\frac{E}{100~\mathrm{keV}})^{\beta }\mathrm{exp}(\beta -\alpha )(\frac{(\alpha-\beta)E_{c}}{100~\mathrm{keV}})^{\alpha-\beta }, E > (\alpha-\beta) E_{c}
\end{array}\right.
\end{eqnarray}
where $A$ is the normalization of the spectrum, and $\alpha$ and $\beta$ are the low and high-energy photon spectral indices, respectively. $E_{\rm p}=(2+\alpha)E_{\rm c}$ is the peak energy, and $E_{\rm c}$ is the cut-off energy.

(2). The CPL model is expressed as,
\begin{eqnarray}
N_{\rm CPL}(E) = A\cdot (\frac{E}{100~\mathrm{keV}})^{\alpha} \mathrm{exp}(-\frac{E}{E_{c}}).
\end{eqnarray}

(3). BB emission of quasi-thermal component \citep{2010ApJ...709L.172R} can be expressed as the Planck function,
\begin{equation}
N_{\textrm{BB}}(E)=A\cdot \frac{E^{2}}{\exp [E / k T]-1}
\end{equation}
where $k$ and $T$ are Boltzmann constant and temperature, respectively.

(4). A PL component is written as
\begin{equation}
N_{\textrm{PL}}(E)=A\cdot (\frac{E}{100~\mathrm{keV}}) ^{\Gamma}
\end{equation}
where $\Gamma$ is the photon index of the spectrum.

Both time-integrated and time-resolved spectra of our sample are extracted from the TTE data. We adopt a Bayesian analysis of the package named the Multi-Mission Maximum Likelihood framework (3ML) to extract the spectrum and then perform the spectral fits \citep{2017ifs..confE.130V}. First, we extract the raw photon count spectrum from the TTE data. For the background deduction of GRB, we select two time intervals before and after the prompt emission of GRBs as the background, and fit the light curve of the background with order 0-4 polynomials as the background emission of GRB to form new observed data. The spectral energy ranges from 8 to 900 keV for the NaI detectors and 250 keV-40 MeV for the BGO detectors, respectively. In order to avoid the K-edge at 33.17 keV caused by the instrument itself, we exclude the energy range from 30 to 40 keV. Then, we invoke the fitting model, which is used to convolve with the standard response files (RSP or RSP2) and generate a predicted count spectrum to compare with the observed data. Maximum-likelihood-based statistics called PGstat (Pgstat $=$ -2ln$L$) are used, with a Poisson likelihood for observation \citep{1979ApJ...228..939C} and Gaussian likelihood for background, respectively. Finally, we adopt the 3ML to perform Bayesian analysis and obtain the best-fit parameters with uncertainties.

Bayesian analysis is indeed to be conducted with the consideration of prior assumptions. In this task, the prior distributions are defined as follows: (1) for the Band model, the normalization ($A$) is assigned a log-normal distribution of prior with $\mu =$ 0 and $\sigma = 2$, where $\mu$ and $\sigma$ are the mean value and standard deviation of the distribution, respectively. The peak energy ($E_{\rm p}$) is also assumed to be a log-normal distribution with $\mu=$2 and $\sigma=$1. The low energy index ($\alpha$) with $\mu =$ -1 and $\sigma =$ 0.5 (lower bound = -1.5, upper bound = 1) and the high-energy index ($\beta$) with $\mu =$ -2 and $\sigma =$ 0.5 (lower bound = -5, upper bound = -1.8) are assumed to be truncated-Gaussian distributions. (2) For the CPL model, it is similar to that of the Band model with log-normal or truncated-Gaussian distribution; the normalization ($A$) is assigned a log-normal distribution of prior with $\mu =$ 0 and $\sigma = 2$. The cut energy ($E_{\rm c}$) with $\mu =$ 2 and $\sigma =$ 1 is also assumed to be a log-normal distribution. The spectrum index ($\alpha$) with $\mu =$ -1 and $\sigma =$ 0.5 (lower bound = -5, upper bound = 1) is assumed to be a truncated-Gaussian distribution. (3) For the PL model, the spectrum index $\Gamma$ is assumed to follow a truncated-Gaussian prior with $\mu =$ -2, $\sigma =$ 1 (lower bound = -5, upper bound = 1), and the normalization ($A$) is assigned a log-uniform prior with $A$ distributed within the range of $10^{-7}$-10. (4) For the BB model, the normalization ($A$) and the black body temperature ($kT$) are assumed to be log-uniform  priors with $10^{-10}$-$10^{3}$ and 1 to 1000, respectively. The posterior distribution is derived by combining the prior distribution with the information of the sample, which is obtained from the data set. As more information about the sample (i.e., larger data set) becomes available, the less influence the prior distribution has on the posterior distribution \citep{2020ApJ...894..100L}.

Accurately calculating the posterior distribution in Bayesian statistical inference presents a significant challenge, which prevents an analytic solution from being obtained through Bayesian posterior sampling, except in the simplest posterior cases. Therefore, stochastic sampling techniques, such as Markov chain Monte Carlo (MCMC, $\mathit{emcee}$, \citealt{2010CAMCS...5...65G}), or nested ($\mathit{MultiNest}$, \citealt{2009MNRAS.398.1601F,2013AIPC.1553..106F}) sampling methods, must be considered. In this paper, we apply the $\mathit{emcee}$ method to sample the posterior and obtain the best-fit parameters \footnote{Here, we also adopt $\mathit{MultiNest}$ method to compare with using $\mathit{emcee}$, and find that the results obtained by both $\mathit{emcee}$ and $\mathit{MultiNest}$ methods are pretty similar to each other in the majority of cases.}. The uncertainties of the fit are typically provided at the 68$\%$ Bayesian confidence level and calculated using the last 75$\%$ of the MCMC chain with 10000 iterations. This is to ensure that the uncertainties are estimated from a sufficiently converged portion of the chain.

In order to find the best model of the fitting, we adopt the Bayesian information criterion (BIC) to judge. The definition of BIC can be expressed as BIC=-2ln $L+k\cdot$ ln($n$), where $L$ is the maximum value of the likelihood function of the estimated model. $k$ and $n$ are the number of model parameters and data points, respectively. In fact, the value of BIC itself does not directly represent the absolute performance or predictive power of a model, but rather provides a relative metric for selecting among multiple models. BIC is a criterion to evaluate the best model fit among a finite set of models, and the model with the lowest BIC is preferred \citep{Neath2012}. \citet{2016A&A...588A.135Y} found that the CPL model is the best one to fit time-resolved spectrum for a majority of GRBs in Fermi/GBM data. Therefore, in our analysis, we adopt the CPL model as the based model for the fits.

On the other hand, we define $\Delta$BIC = BIC ($\mathit{i}$)-BIC (CPL) to find out the best model, where $\mathit{i}$ = Band, BB, CPL+BB, Band+BB and PL+BB. If $0<\rm \Delta BIC<2$, the evidence against the model with the higher BIC is not worth more than a bare mention; if $2<\rm \Delta BIC<6$, the evidence against the model with the higher BIC is positive; if $6<\rm \Delta BIC<10$, the evidence against the model with the higher BIC is strong; if $10<\rm \Delta BIC$, the evidence against the model with the higher BIC is very strong. In other words, if $\rm \Delta BIC$ is a negative value, it means that the model we adopted is better than the CPL model.

\section{Results}{\label{s3}}
\subsection{Time-integrated spectra fits}
\begin{figure}[htb]
\centering
 \includegraphics [angle=0,scale=0.4] {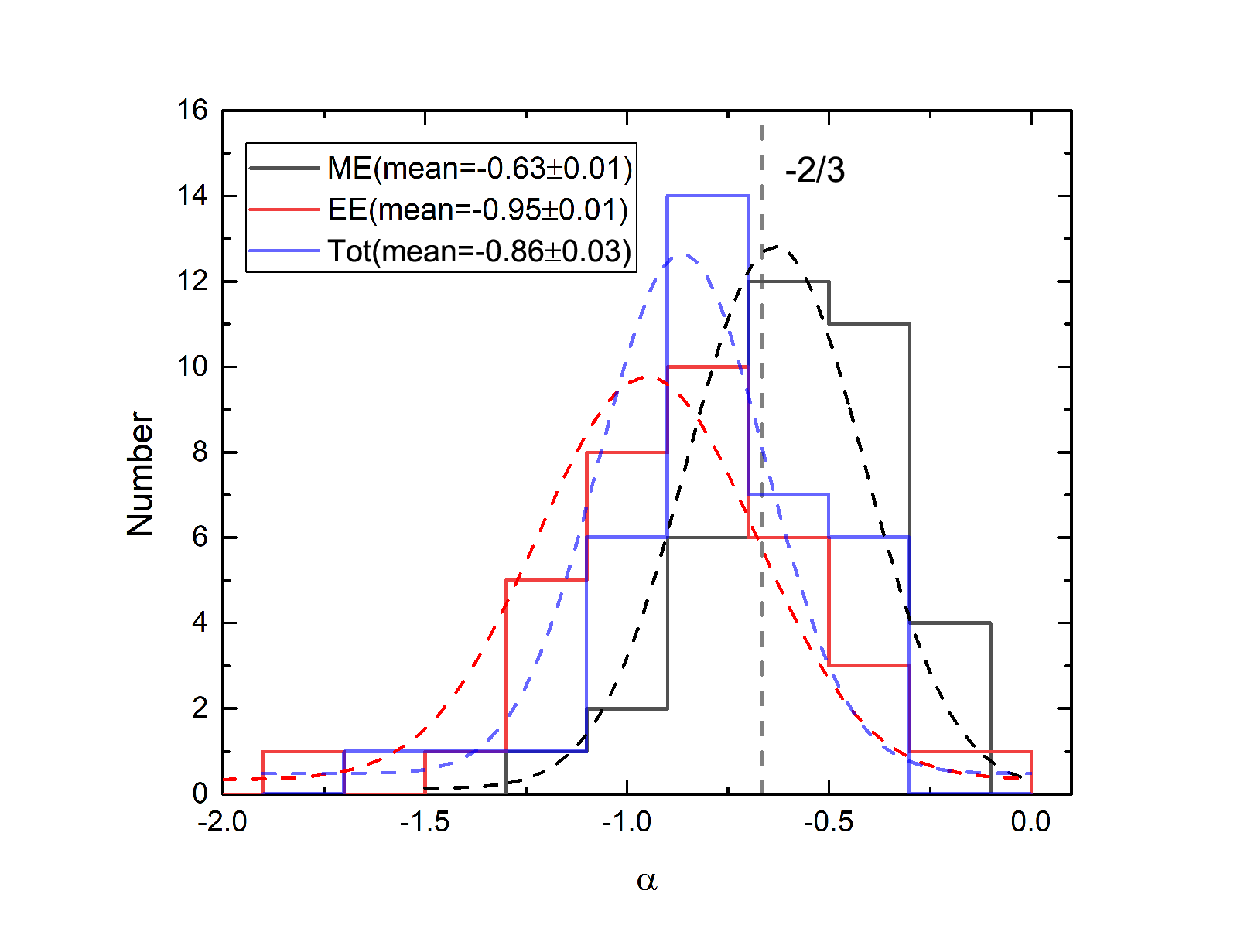}
 \caption{The distributions of $\alpha$ and Gaussian fitting for ME (black solid line), EE  (red dashed line), and whole burst ( blue dotted line). The gray dashed line is the death line of synchrotron emission.}
 \label{fig:cpl-band}
\end{figure}

First, we perform a preliminary analysis of the 36 GRBs from the perspective of time-integrated spectra. For each GRB, it includes the ME and EE components that are also used to do the spectral fitting, respectively. Then, we adopt the Band or CPL model to do the time-integrated spectral fits for the whole burst, ME, and EE components, respectively. Figure \ref{fig:cpl-band} shows the distributions of $\alpha$ of spectral fitting with CPL or Band function, and the fitting results are presented in Table \ref{Tablel}. We find that there exists a significant difference in the distribution of $\alpha$ between the ME and EE components. The distribution of $\alpha$ of ME peaks around -0.63, which is systematically harder than that of EE peaks around -0.95. For the whole burst of our sample, the distribution of $\alpha$ peaks at -0.86 $\pm$ 0.03. It means that the $\alpha$ values of a fraction of our sample at least exceed the death line ($\alpha=-2/3$) of the requirement of synchrotron radiation.

Second, based on the results of $\alpha$ distribution above, it is found that there are 20 out of 36 GRBs for which $\alpha$ values exceed -2/3 within either their ME or EE phases. We divide them into three categories based on where the $\alpha$ exceeds the death line of synchrotron radiation: (I) The $\alpha$ value in the ME phase surpasses -2/3, while the EE phase does not. There are altogether 12 robust cases (e.g., GRBs 090720B, 100829A, 111012B, 120119B, 150127A, 150510A, 150702A, 160721A, 170115B, 170626A, 210202A, and 210520A). (II) The $\alpha$ values exceed -2/3 in both ME and EE phases, and six GRBs can be grouped into this category (e.g., GRBs 080807, 081215, 090929A, 120304B, 141229A, and 221112A). (III) The $\alpha$ value exceeds -2/3 in the EE phase but does not in the ME phase, and it only includes two cases, e.g., GRBs 161218B and 220903A. From the physical point of view, it is very difficult to explain such large $\alpha$ values by adopting synchrotron emission for those 20 GRBs. So we focus on studying those 20 GRBs for which $\alpha$ values exceed the death line of synchrotron emission and try to investigate the jet composition of GRB. However, the information gleaned from time-integrated spectra is limited, and it is unclear whether the observed time-integrated spectra of GRB are intrinsic or caused by the superposition of different types of spectra. So, it is necessary to analyze the time-resolved spectra for those 20 GRBs.

\subsection{Time-resolved spectra fits}
We perform time-resolved spectra analysis of the above 20 GRBs in our sample, and the TTE data is rebinned using the of signal to noise ratio (S/N) method, which is calculated by $\mathit{Significance}$ \citep{2018ApJS..236...17V}. In our analysis, three criteria are required to perform the time-resolved spectra analysis of those 20 GRBs: (1) S/N = 30  is adopted for each time bin of the TTE light curve; (2)  each time bin has a duration of more than 0.03 s; (3) in order to show the evolution behaviors of the spectra, one needs to include at least five time bins for the ME, EE, or whole burst. Following the above criterion, there are only 6 out of 20 GRBs that can be used for the time-resolved spectra analysis, e.g., GRBs 081215, 100829A, 150510A, 161218B, 170115B, and 170626A. In the following, we only focus on analyzing the time-resolved spectra of those six bright GRBs.

\subsubsection{GRB 081215}
\begin{figure}
\centering
 \includegraphics [angle=0,scale=0.3] {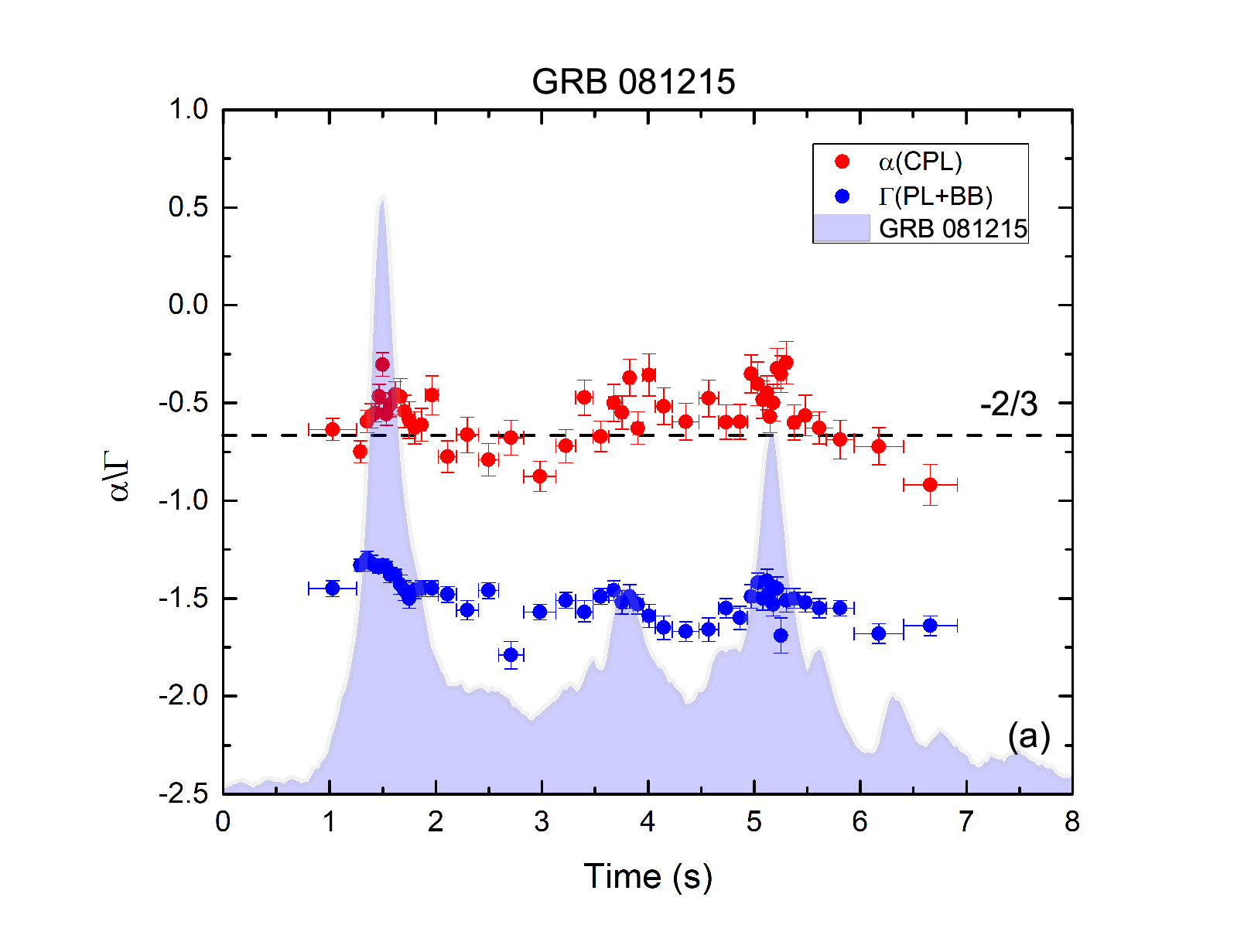}
\includegraphics [angle=0,scale=0.3] {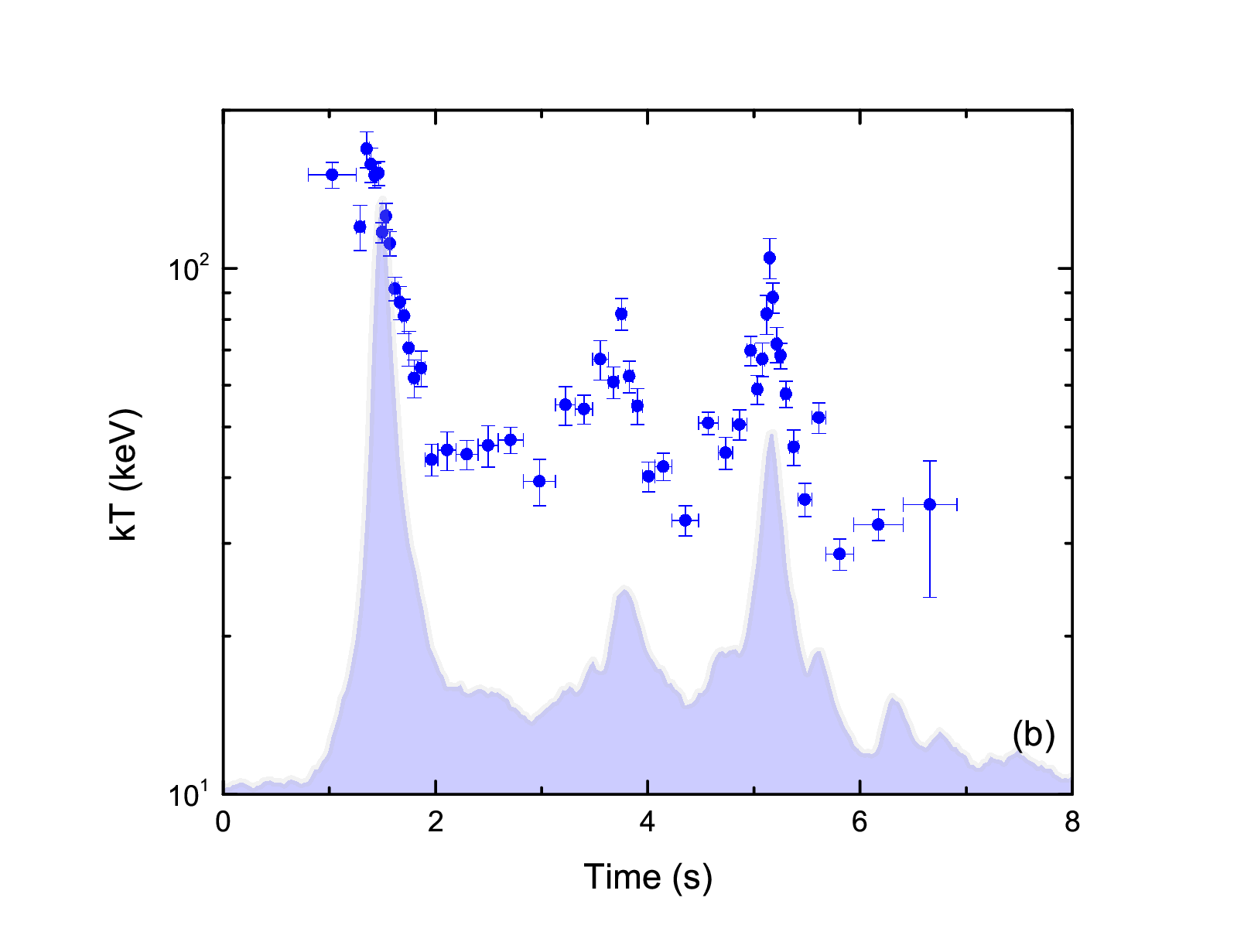}
 \includegraphics [angle=0,scale=0.3] {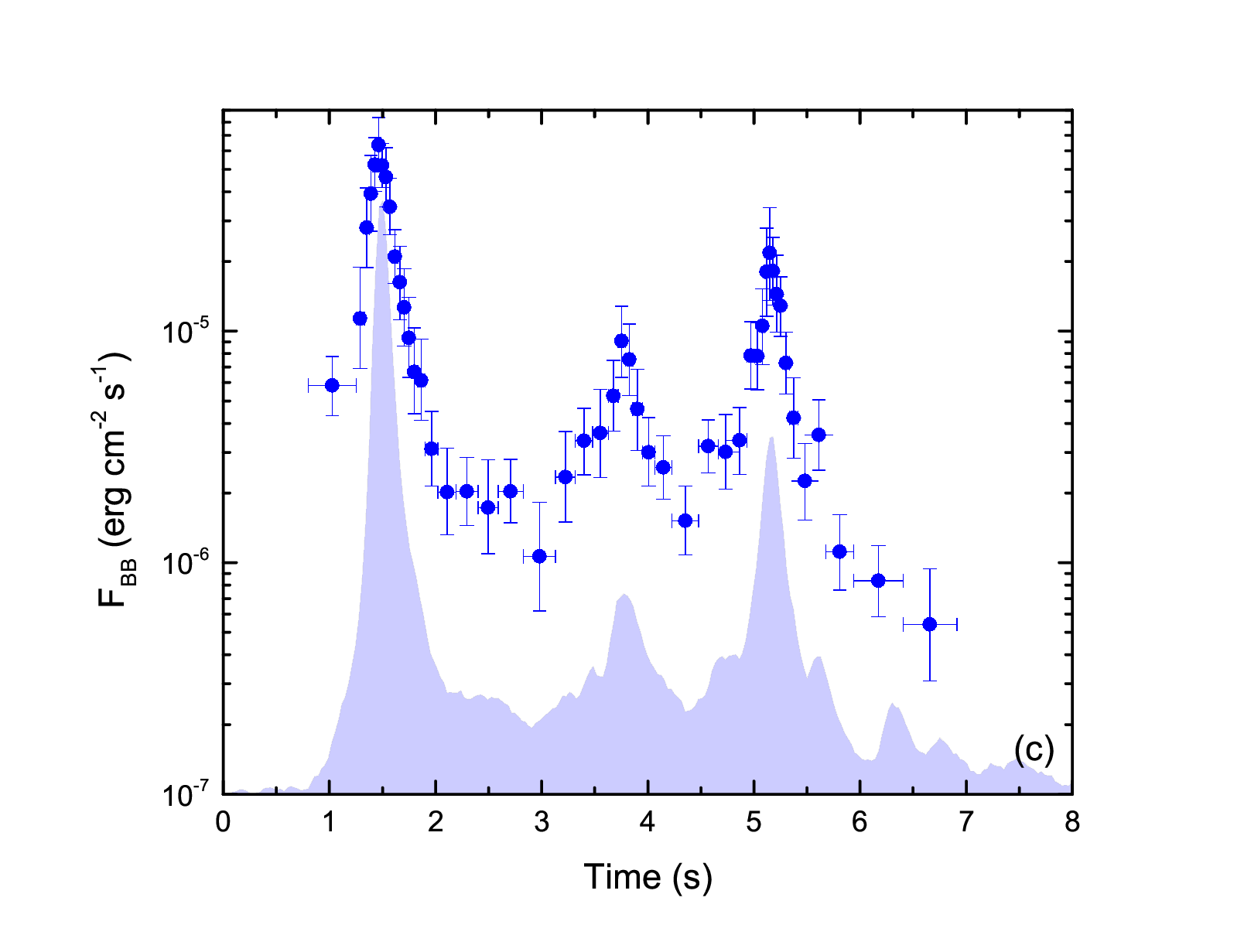}
\includegraphics [angle=0,scale=0.3] {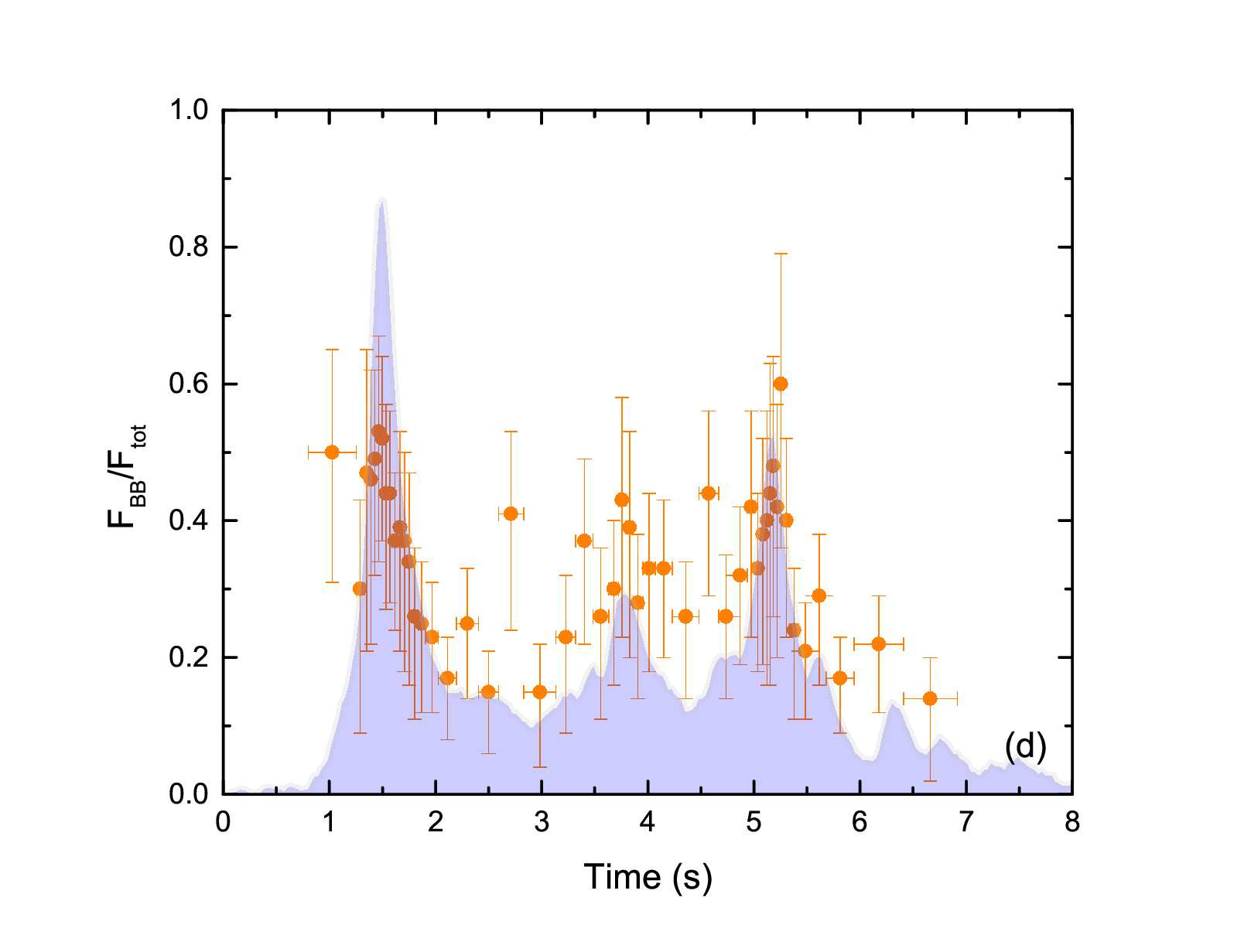}
 \caption{Temporal evolution of $\alpha$ and $\Gamma$ (a), $kT$ (b), $F_{BB}$ (c), and $F_{BB}/F_{tot}$ (d) of GRB 081215. The dashed line is the limiting value of $\alpha$ = -2/3.}
     \label{fig:GRB 081215}
\end{figure}

The Band model is the best one to fit the time-integrated spectra of GRB 081215 for both ME and EE components, and it belongs to category (II) with $\alpha$ values exceeding $-2/3$ in both ME and EE phases. Based on the descriptions in Section 3.2, we first extract the time-resolved spectra of GRB 081215 to adopt the CPL model \citep{2020MNRAS.492.3622L}, and we find that the low-energy index $\alpha$ of CPL model continues to exceed $-2/3$ for most time slices of the time-resolved spectra (see Figure \ref{fig:GRB 081215}(a) and Table \ref{Table3}). Then, we adopt Band, CPL, BB, PL$+$BB, CPL$+$BB, and Band$+$BB to do the time-resolved spectral fits and find the best fitting model by comparing the $\Delta$BIC for different models. It is found that the PL$+$BB model improves the fitting results with well-converged parameters and minimum BIC values for each case (see Table \ref{Table2}). Note: 'Non-Con' means that the posterior parameters of fits are not constrained. This finding is in agreement with the validation study conducted by \citet{2020MNRAS.492.3622L}. The fitting results are shown in Table \ref{Table3}, and we adopt the ratio between the residual sum of squares (RSS) and the degrees of freedom (RSS/dof) to reflect the constrained condition of the residuals.

Moreover, we also calculate the flux of the thermal component ($F_{\rm BB}$) and the flux ratio between the thermal component the and total flux ($F_{\rm tot}$). Figure \ref{fig:GRB 081215} shows the temporal evolution of $\alpha$ or $\Gamma$ (a), $kT$ (b), $F_{\rm BB}$ (c), and $F_{\rm BB}/F_{\rm tot}$ (d). Here, the flux calculation ranges from 1 to 10000 keV. We find that the low-energy index $\alpha$ obtained from the CPL model exceeds the synchrotron emission death line. However, the values of $\alpha$ in the CPL model decreased for most time slices when we adopt the PL$+$BB model for the fits instead of the CPL model; this is, the $\alpha$ in the CPL model is approximately equal to $\Gamma$ in the power-law model. The power-law index $\Gamma$ seems to track the pulses of prompt emission in both the ME and EE phases. On the other hand, the temporal evolution of $kT$ and $F_{BB}$ also seems to track the pulses of prompt emission, but $F_{\rm BB}/F_{\rm tot}$ is not as obvious.

\subsubsection{GRB 100829A}
\begin{figure}
\centering
 \includegraphics [angle=0,scale=0.3] {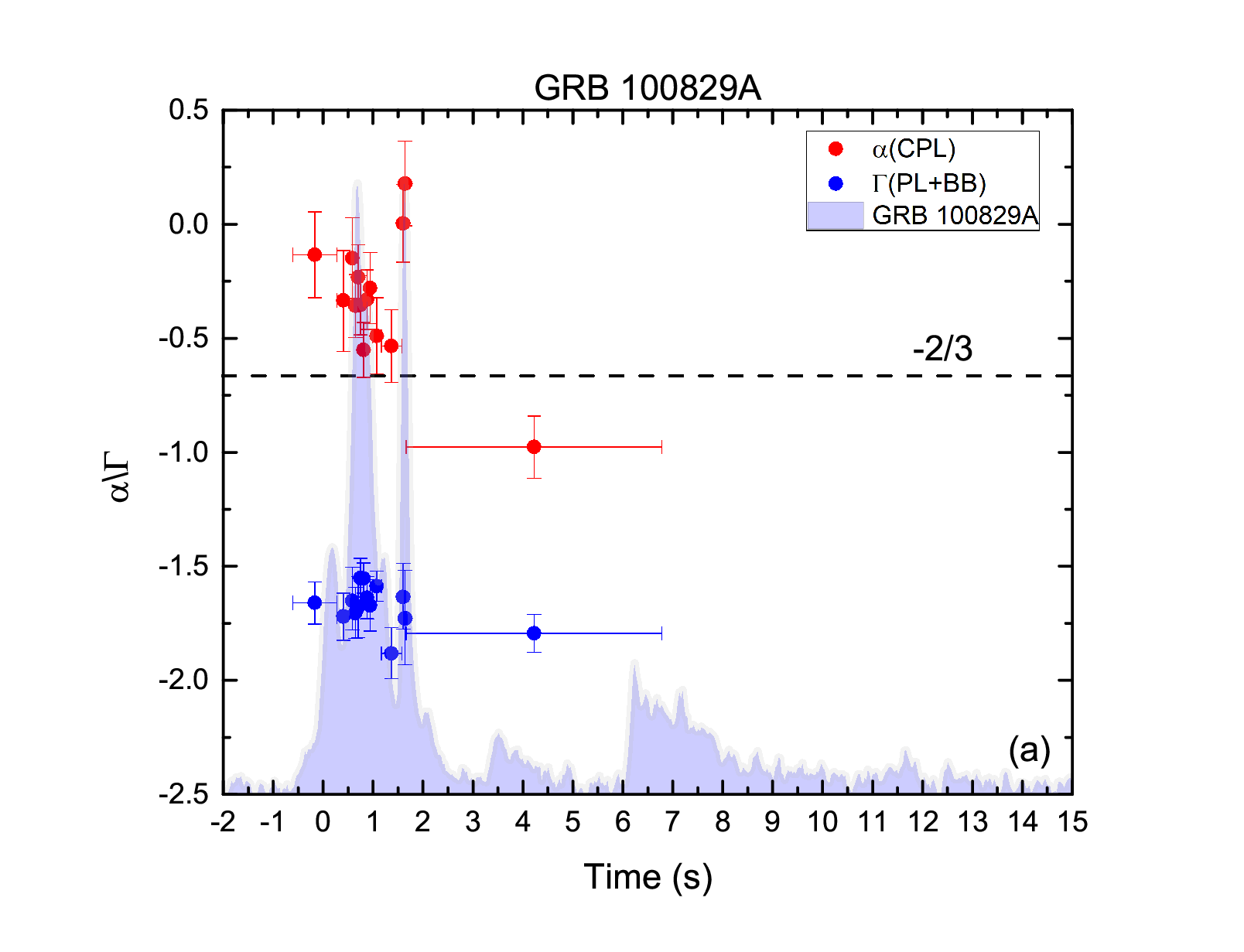}
\includegraphics [angle=0,scale=0.3] {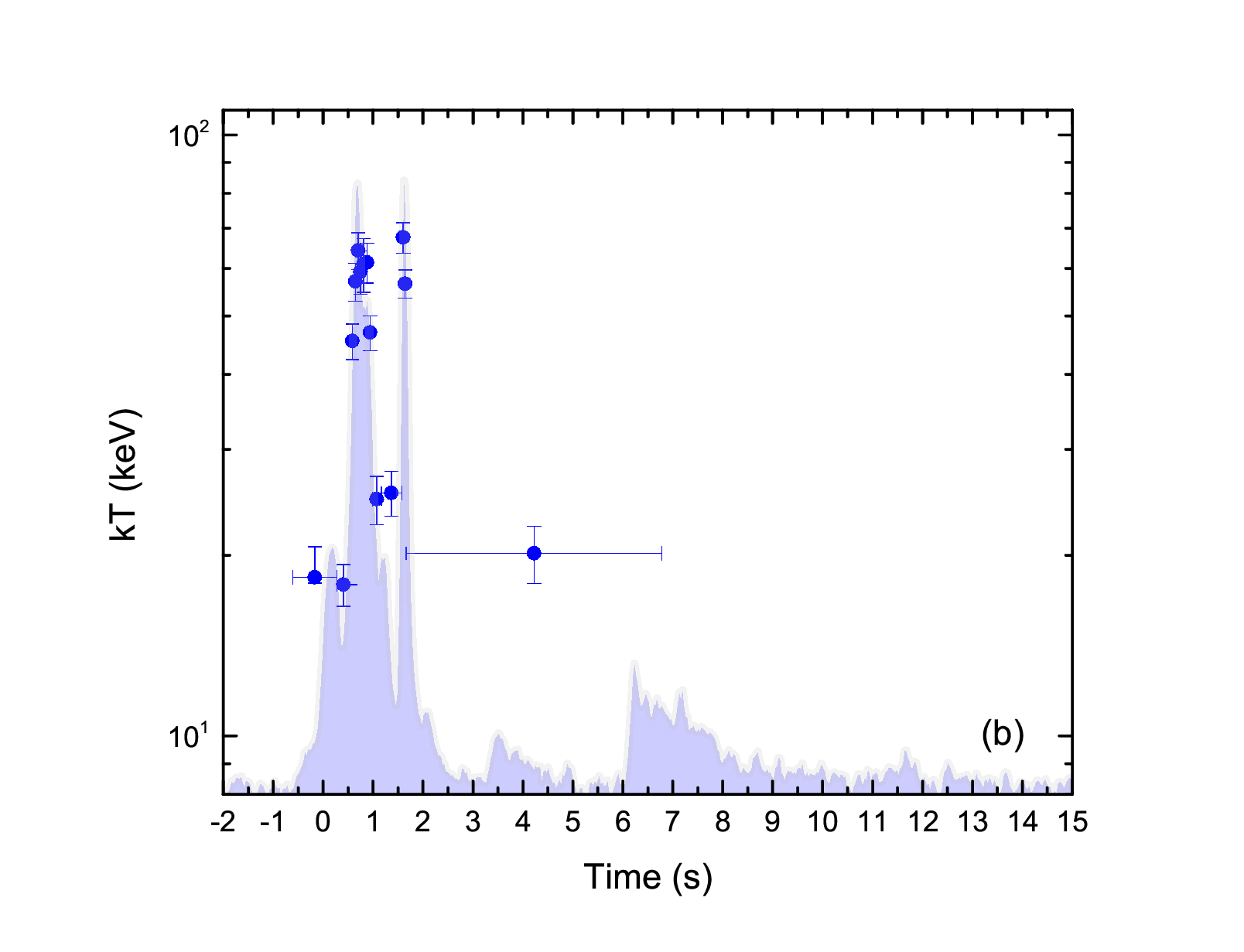}
 \includegraphics [angle=0,scale=0.3] {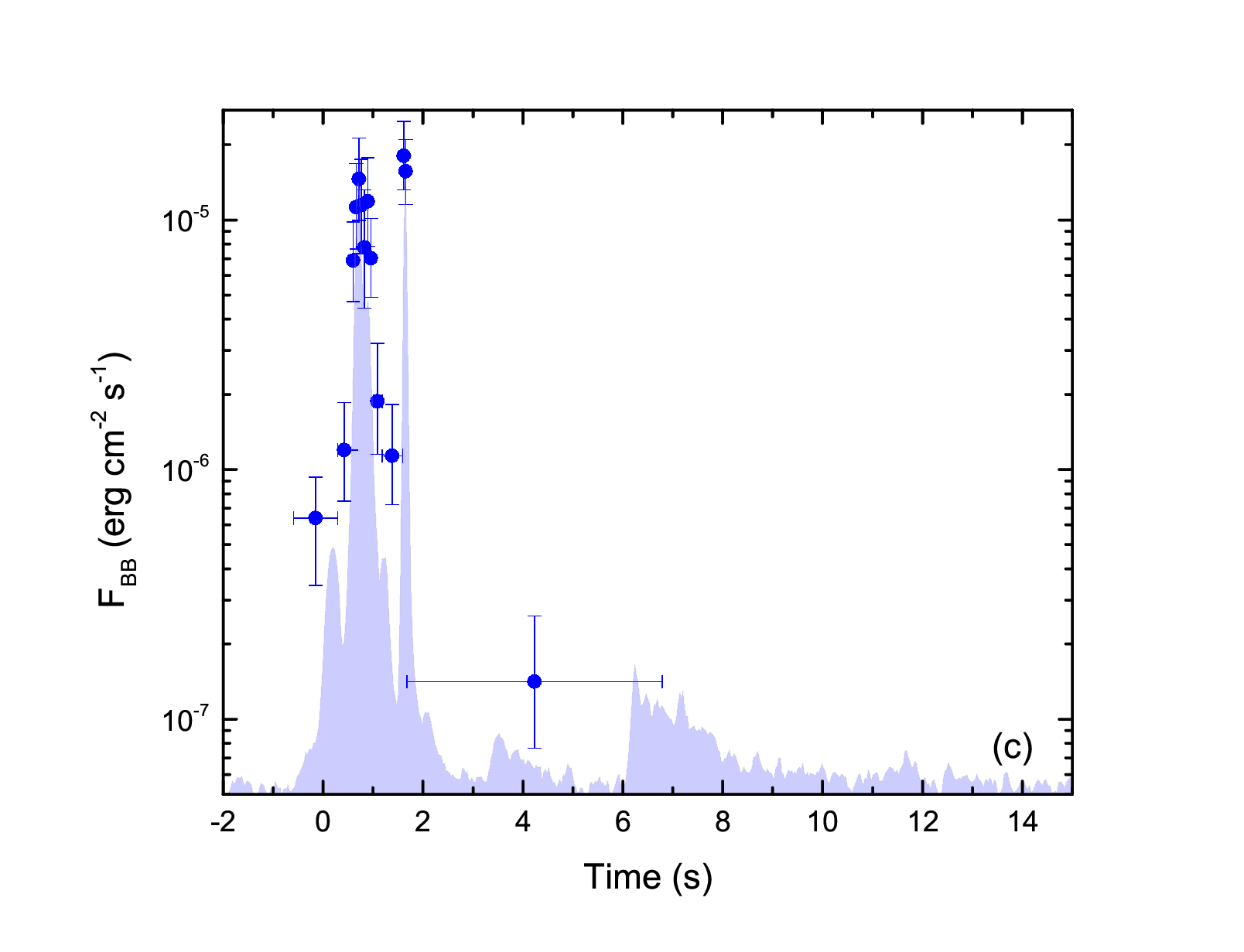}
\includegraphics [angle=0,scale=0.3] {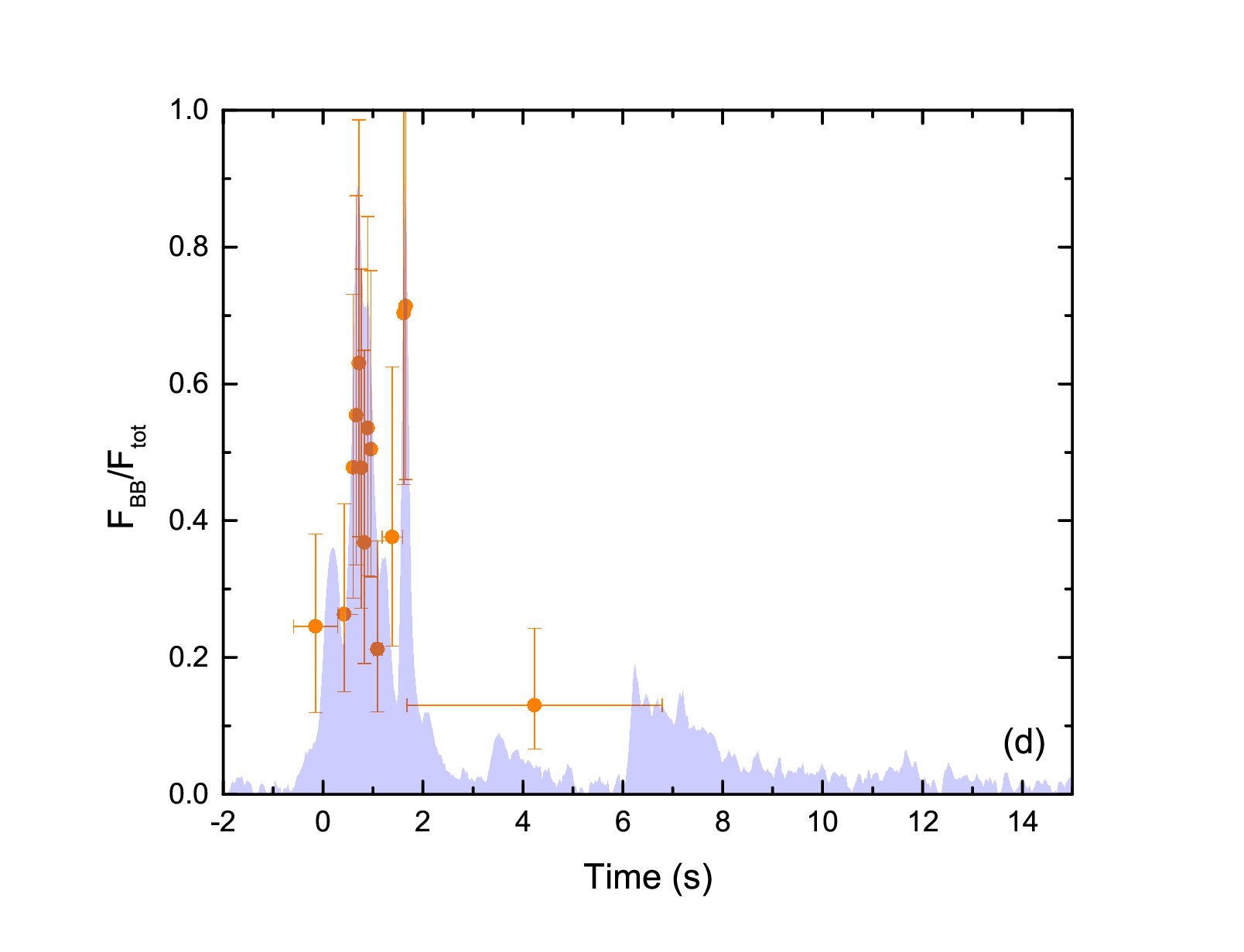}
 \caption{Similar to Figure 3, but for GRB 100829A.}
\label{fig:GRB 100829A}
\end{figure}

The time-integrated spectrum of GRB 100829A belongs to category (I) for which $\alpha$ values exceed -2/3 in the ME phase, but the EE phase does not. The Band and CPL models are the best ones to fit in the ME and EE components, respectively. Initially, we also extract the time-resolved spectra of GRB 100829A to adopt the CPL model, which is similar to the approach taken for GRB 081215. We find that the low-energy index $\alpha$ of the CPL model continues to exceed $-2/3$ for all time slices of the time-resolved spectra in the ME phase (see Figure \ref{fig:GRB 100829A}(a) and Table \ref{Table5}). Then, we compare several models (e.g., Band, CPL, BB, PL$+$BB, CPL$+$BB, and Band$+$BB) based on the values of $\Delta$BIC for different models. It is found that the PL$+$BB model is the best one for the time-resolved spectra of GRB 100829A (see Table \ref{Table4}). The fitting results are presented in Table \ref{Table5}.

Similar to the Figure \ref{fig:GRB 081215} of GRB 081215, Figure \ref{fig:GRB 100829A} shows the temporal evolution of $\alpha$ or $\Gamma$ (a), $kT$ (b), $F_{\rm BB}$ (c), and $F_{\rm BB}/F_{\rm tot}$ (d). We find the $\Gamma$ obtained from the power-law model does not show significant evolution. However, the evolution of $kT$, $F_{\rm BB}$, and $F_{\rm BB}/F_{\rm tot}$ exhibit behavior of tracks the variability of the light curve in GRB 100829A.

\subsubsection{GRB 161218B}
\begin{figure}
\centering
 \includegraphics [angle=0,scale=0.3] {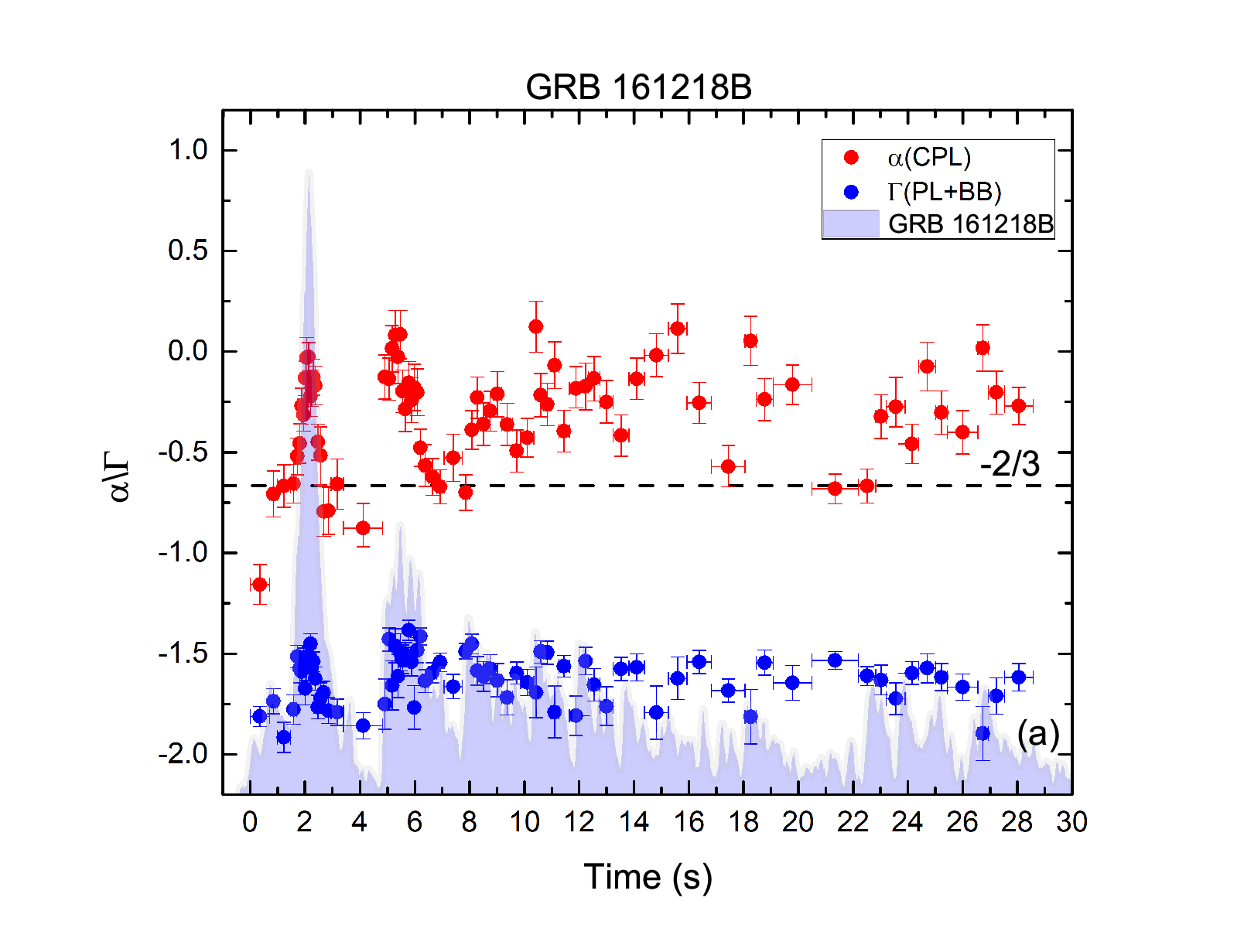}
\includegraphics [angle=0,scale=0.3] {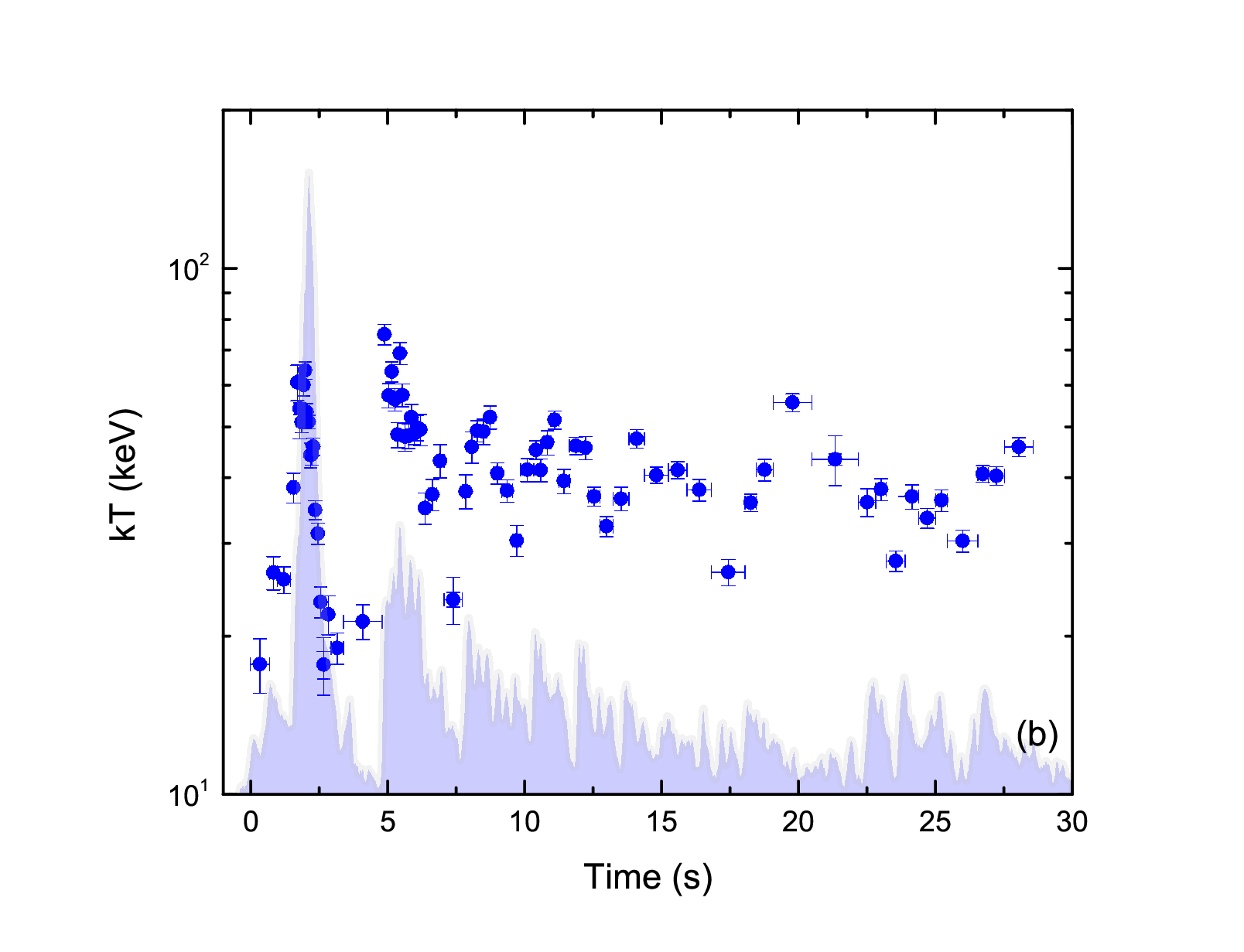}
 \includegraphics [angle=0,scale=0.3] {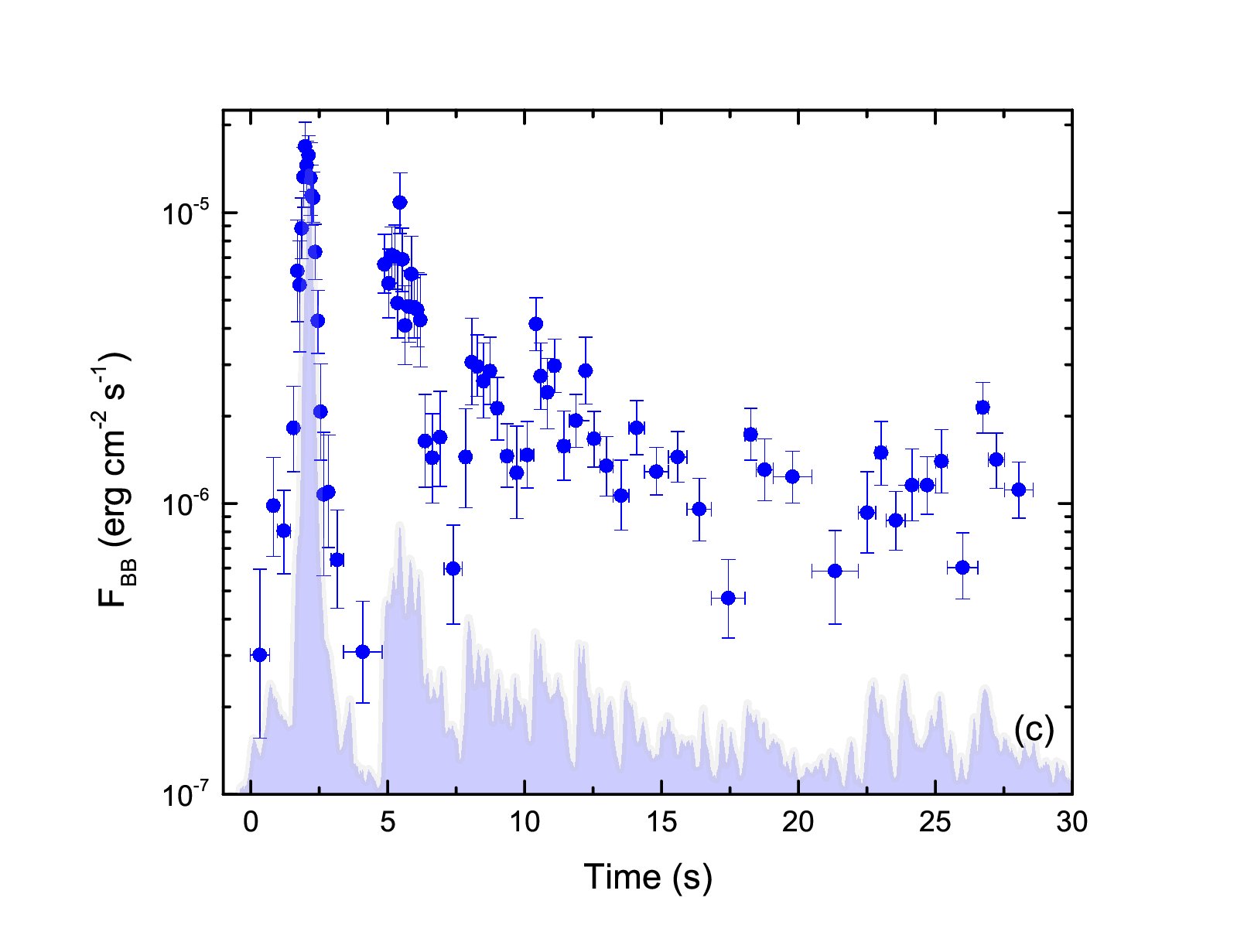}
\includegraphics [angle=0,scale=0.3] {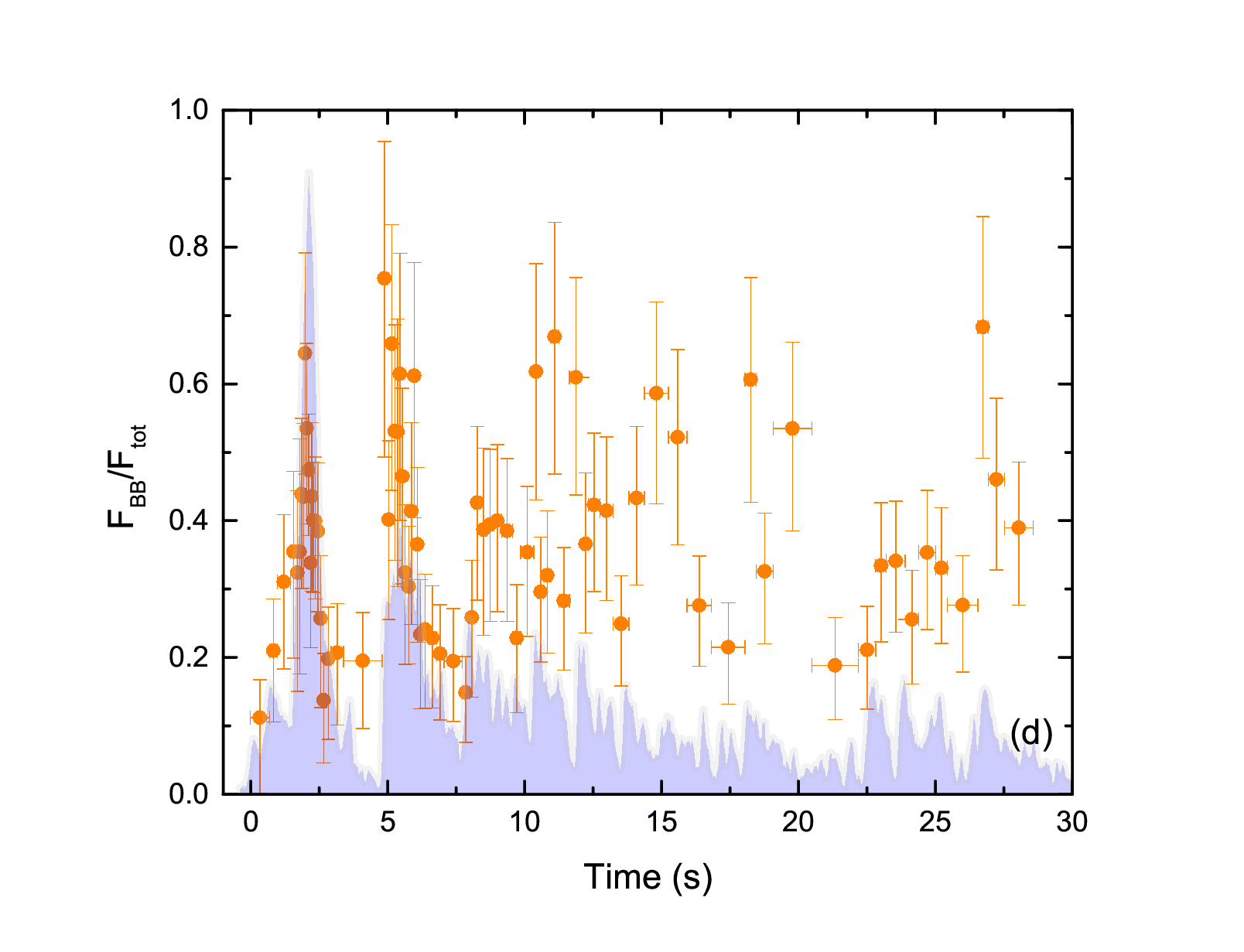}
 \caption{Similar to Figure 3, but for GRB 161218B.}
\label{fig:GRB 161218B}
\end{figure}

The time-integrated spectrum of GRB 161218B belongs to category (III) for which $\alpha$ values exceed -2/3 in the EE phase, but the ME phase does not. The CPL model is the best one to fit the time-integrated spectra in both ME and EE components. By extracting the time-resolved spectra of GRB 161218B, we find that the low-energy index $\alpha$ of the CPL model continues to exceed $-2/3$ for most of time slices of the time-resolved spectra in both ME and EE phases (see Figure \ref{fig:GRB 161218B}(a) and Table \ref{Table7}). Then, we compare several models (e.g., Band, CPL, BB, PL$+$BB, CPL$+$BB, and Band$+$BB) based on the values of $\Delta$BIC for different models. It is found that the PL$+$BB model is the best one for all time slices of GRB 161218B (see Table \ref{Table6}). The fitting results are presented in Table \ref{Table7}.

Figure \ref{fig:GRB 161218B} shows the temporal evolution of $\alpha$ or $\Gamma$ (a), $kT$ (b), $F_{\rm BB}$ (c), and $F_{\rm BB}/F_{\rm tot}$ (d). We find that the $\Gamma$ obtained from the power-law model, $kT$ and $F_{\rm BB}$ obtained from thermal emission, as well as $F_{\rm BB}/F_{\rm tot}$, seem to be track the light curve of prompt emission in GRB 161218B.

\subsubsection{GRB 170626A}
\begin{figure}
\centering
 \includegraphics [angle=0,scale=0.3] {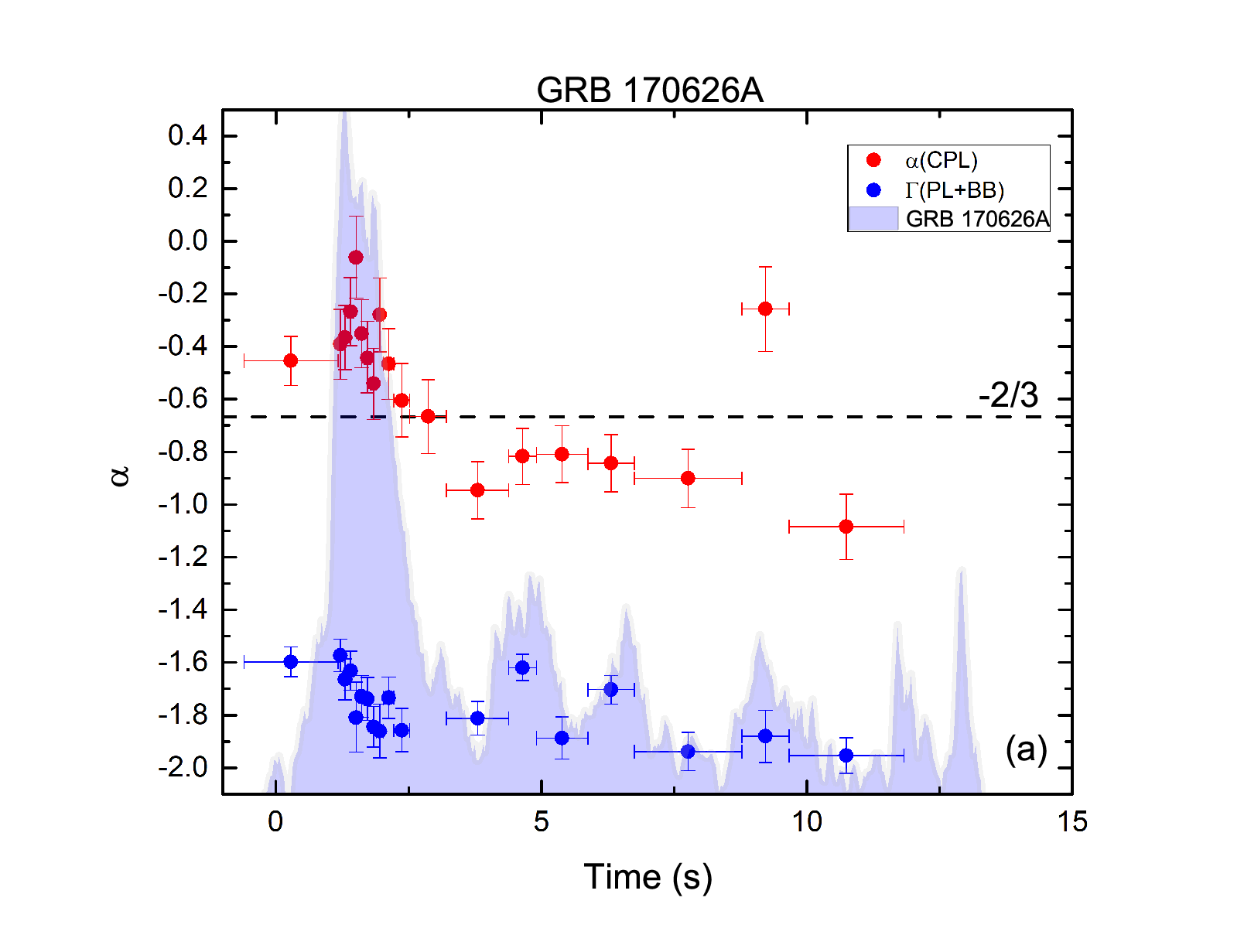}
\includegraphics [angle=0,scale=0.3] {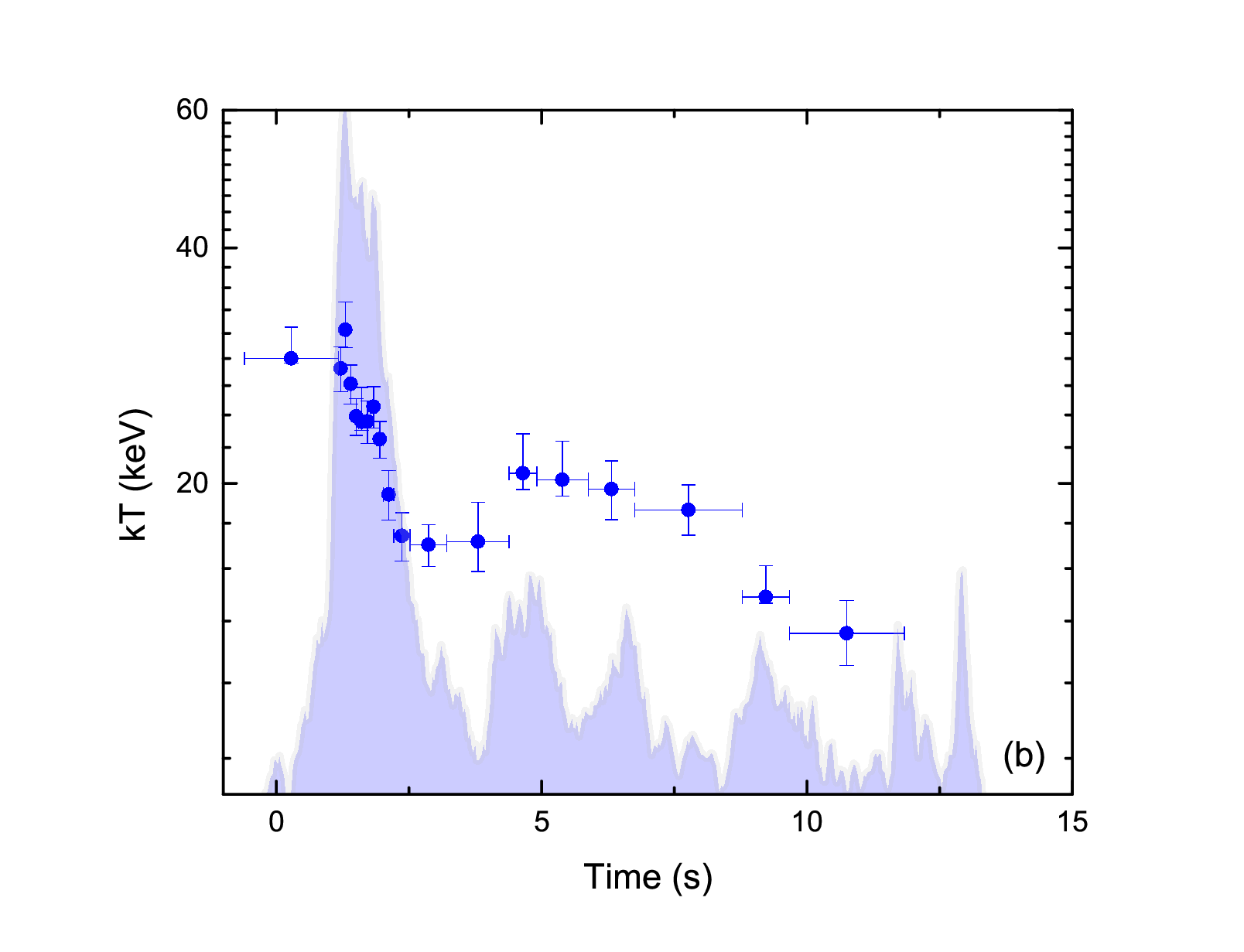}
 \includegraphics [angle=0,scale=0.3] {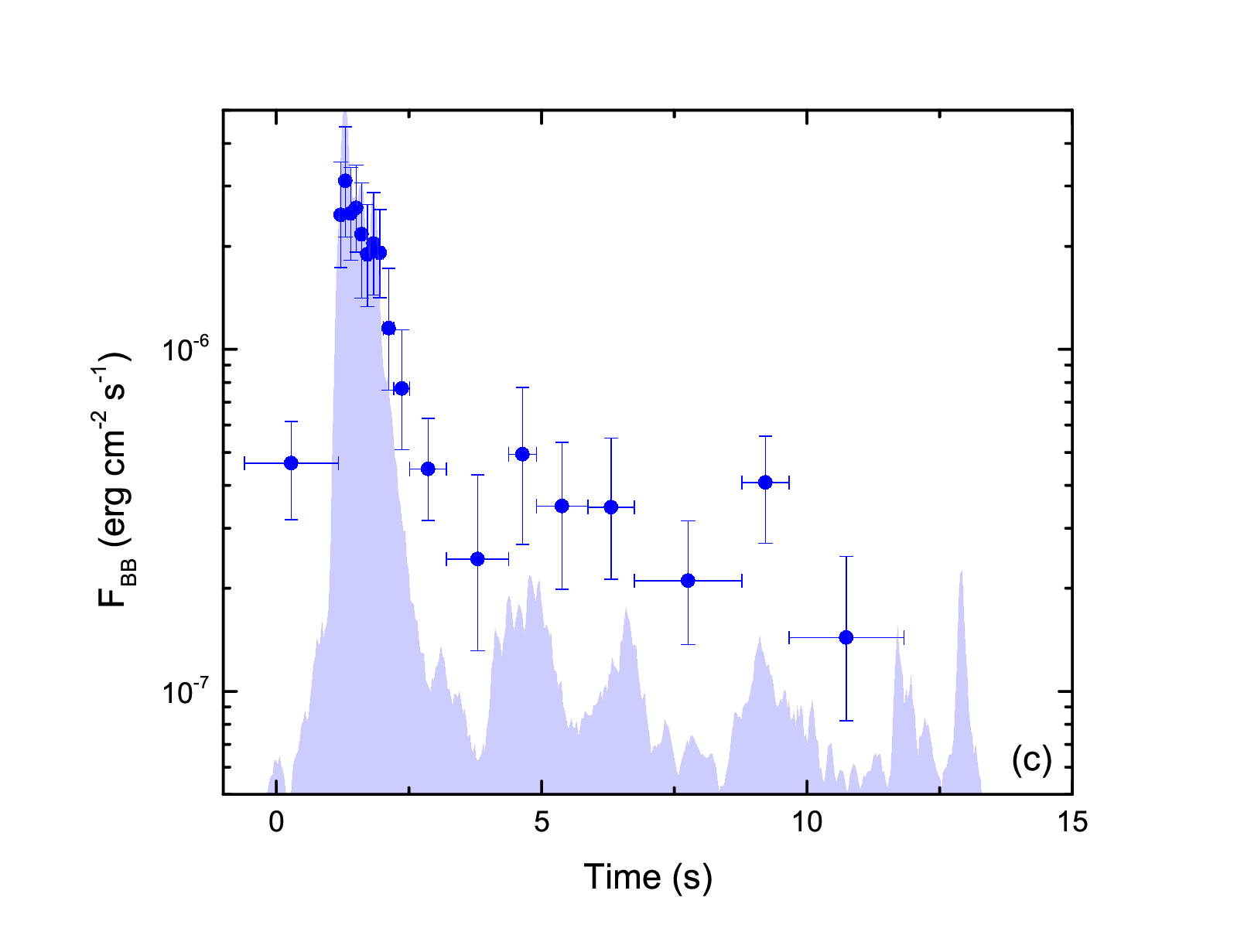}
\includegraphics [angle=0,scale=0.3] {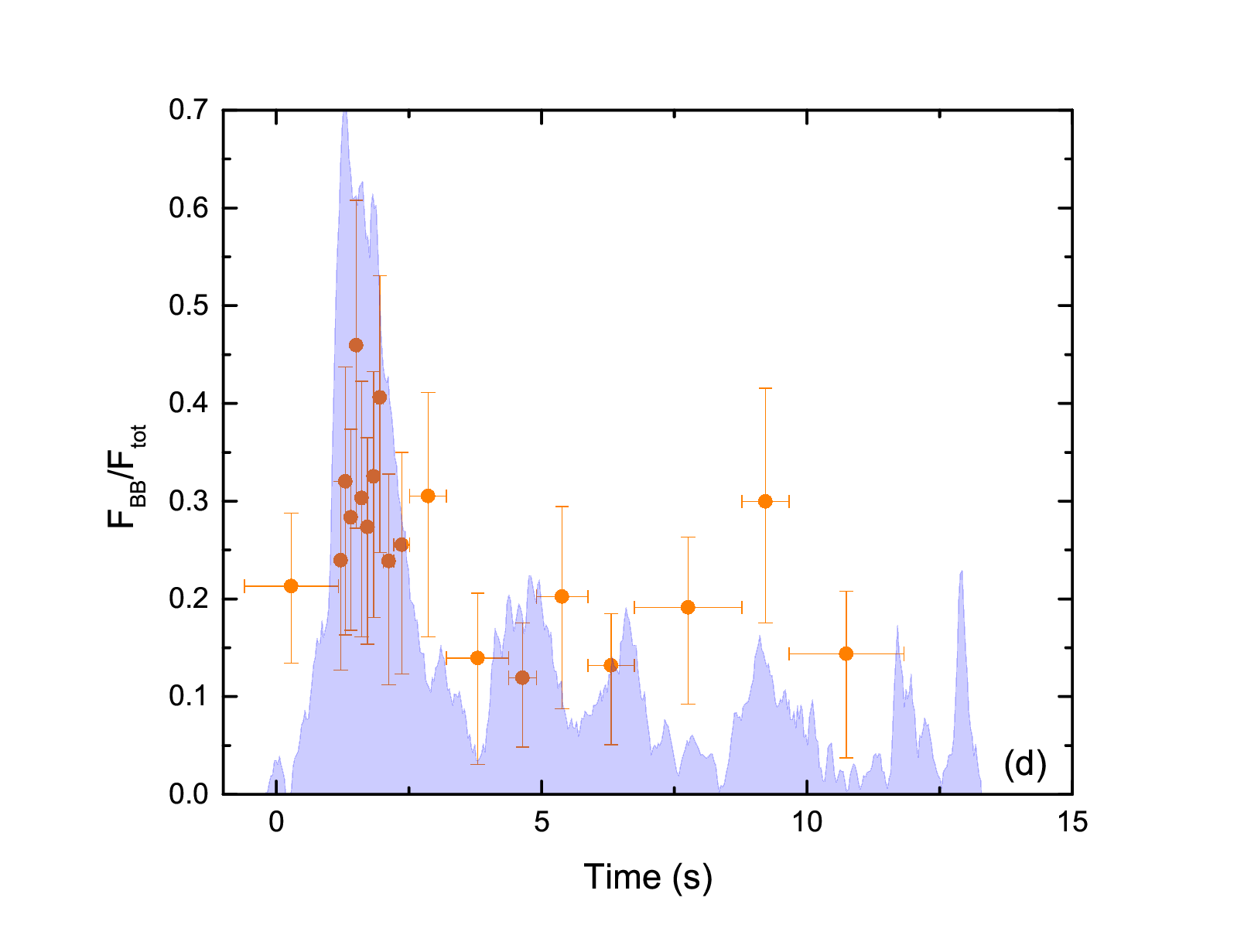}
 \caption{Similar to Figure 3, but for GRB 170626A.}
\label{fig:GRB 170626A}
\end{figure}

The time-integrated spectrum of GRB 170626A belongs to the category (I) for which $\alpha$ values exceed -2/3 in the ME phase, but the EE phase does not, and the Band model is the best one to fit both ME and EE components. The time-resolved spectra of GRB 170626A show that the low-energy index $\alpha$ of the CPL model continues to exceed $-2/3$ for most of the time slices of the time-resolved spectra in the ME phase, but dose not in the EE phase, except one time slice (see Figure \ref{fig:GRB 170626A}(a) and Table \ref{Table9}). We also perform the time-resolved spectra and compare several models based on the values of $\Delta$BIC. It is found that the PL$+$BB model is the best one for all time slices of GRB 170626A (see Table \ref{Table8}). The fitting results are presented in Table \ref{Table9}.

Figure \ref{fig:GRB 170626A} shows the temporal evolution of $\alpha$ or $\Gamma$ (a), $kT$ (b), $F_{\rm BB}$ (c), and $F_{\rm BB}/F_{\rm tot}$ (d). We find that the $\Gamma$ obtained from the power-law model exhibits a hard-to-soft trend in the ME phase but a 'track' trend in the EE phase. $kT$ exhibits an evolution trend from hard to soft in both the ME and EE phases. $F_{\rm BB}$ and $F_{\rm BB}/F_{\rm tot}$ seem to be track the pulses of the light curve in GRB 170626A.

\subsubsection{GRB 150510A and GRB 170115B}
\begin{figure}
\centering
 \includegraphics [angle=0,scale=0.3] {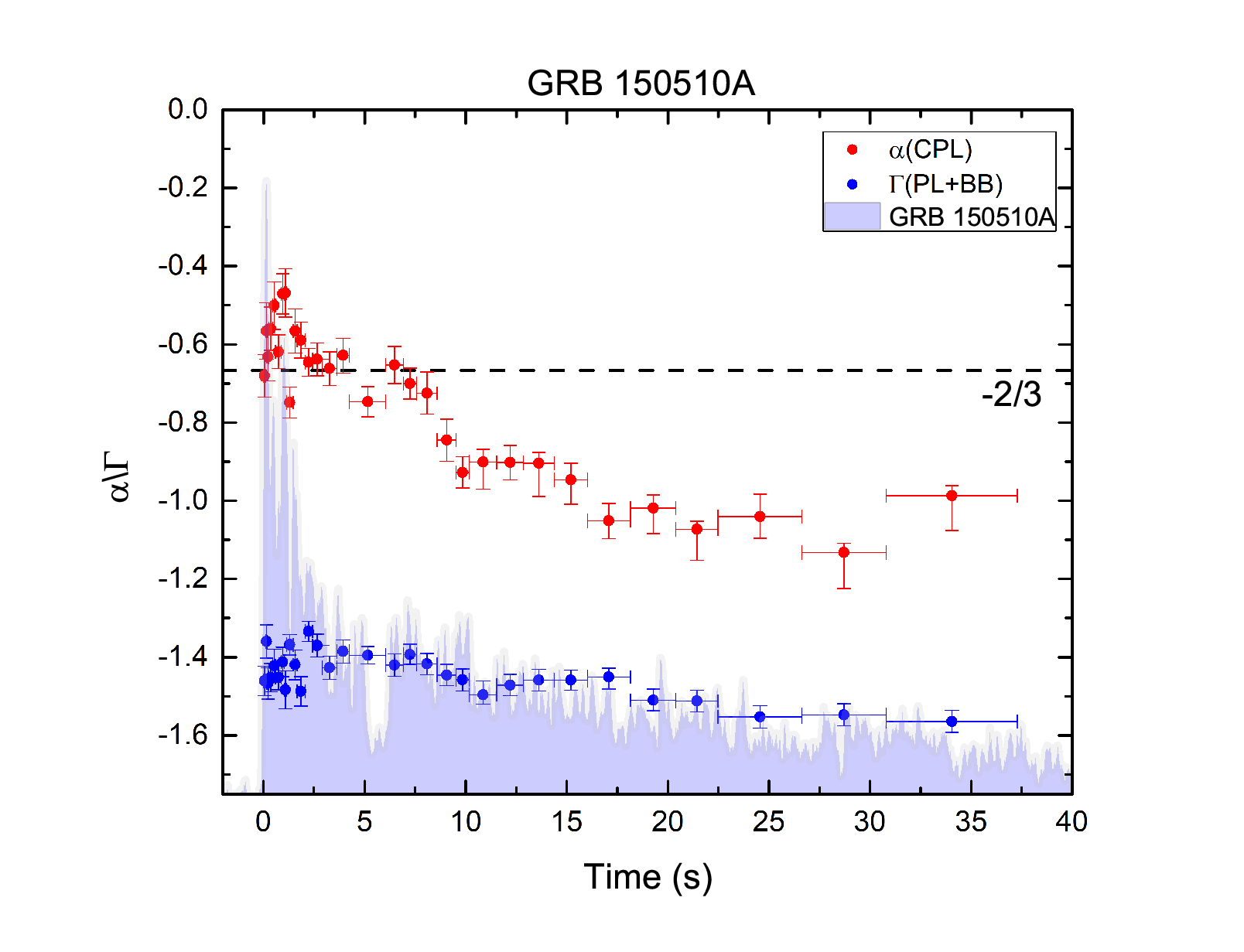}
\includegraphics [angle=0,scale=0.3] {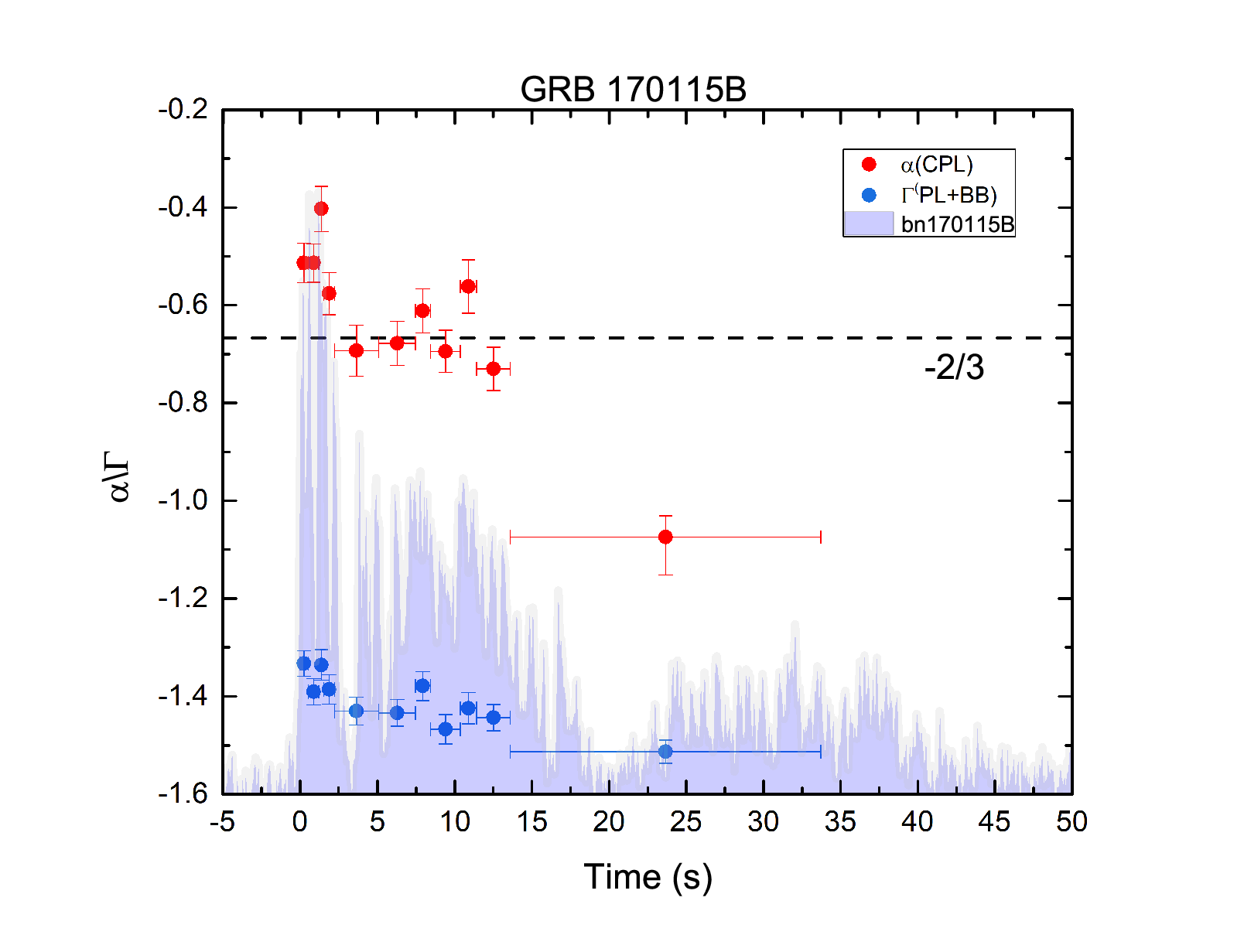}

 \caption{Temporal evolution of $\alpha$ and $\Gamma$ of GRB 150510A and GRB 170115B. The dashed line is the limiting value of $\alpha$ = -2/3.}
     \label{fig:GRB 150510 and 170115}
\end{figure}

The time-integrated spectra of GRB 150510A and GRB 170115B belong to category (I) for which $\alpha$ values exceed -2/3 in the ME phase, but the EE phase does not. For GRB 150510A, it is found that the PL$+$BB model is the best one to fit the EE phase when we extract the time-resolved spectra. However, in the ME phase, there are several time slices of time-resolved spectra that are consistent with the CPL model, even though the $\alpha$ value continues to exceed the death line of synchrotron emission. Moreover, we cannot identify which model (CPL or PL$+$BB) is good enough to describe the time-resolved spectra in seven time slices of GRB 150510A (see Table \ref{Table10}). On the other hand, the PL$+$BB model seems to be consistent with that of time-resolved spectra of GRB 170115B, but there is still uncertainty when compared with the CPL model in three time slices (see Table \ref{Table11}). Due to those uncertainties, we do not discuss its spectral evolution behaviors of GRBs 150510A and 170115B in this paper, and only present $\alpha$ and $\Gamma$ evolutions in Figure \ref{fig:GRB 150510 and 170115}.

\subsection{Statistical correlations of the time-resolved spectra}
Studying the correlation of spectrum parameters is an important approach to understanding the radiation mechanism of GRB itself. In this section, we focus on the time-resolved spectra of those four GRBs (GRBs 081215, 100829A, 161218B, and 170626A) and try to identify some statistical correlations that may provide clues to GRB physics.

\subsubsection{$F_{\rm tot}-\Gamma$}
Previous studies found that a correlation between total flux ($F_{\rm tot}$) and low-energy index $\alpha$ from CPL or Band of nonthermal emission does exist for a large fraction of GRBs \citep{1997ApJ...479L..39C,2002ApJ...565..182L,2014MNRAS.442..419B,2019MNRAS.484.1912R}, and suggested that it is related to the variation of dimensionless energy in the fireball model \citep{2019MNRAS.484.1912R}. Here, we also try to identify whether a correlation exists between $F_{\rm tot}$ and $\Gamma$ from the power-law model for those four GRBs.

\begin{figure}
\centering
 \includegraphics [angle=0,scale=0.3] {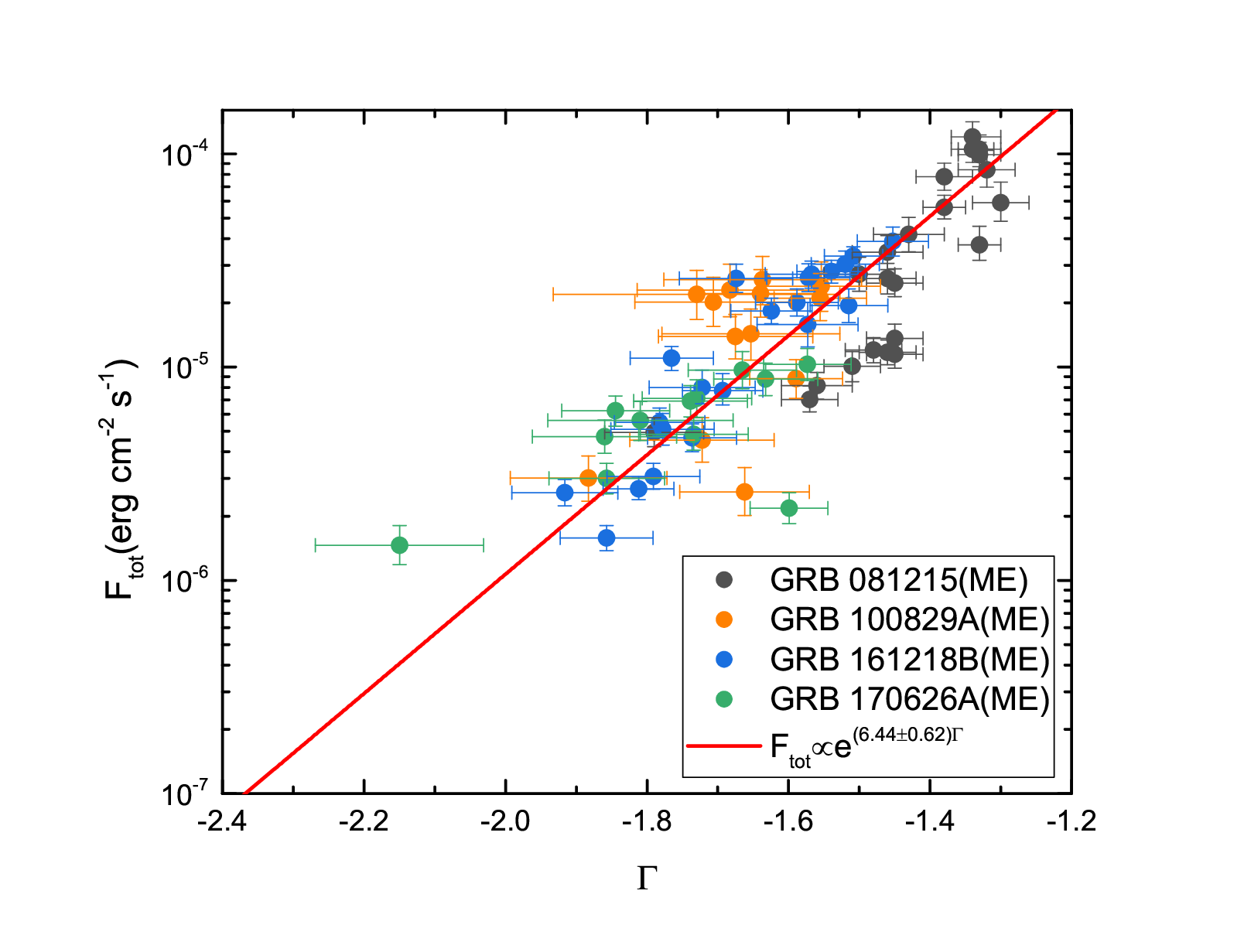}
\includegraphics [angle=0,scale=0.3] {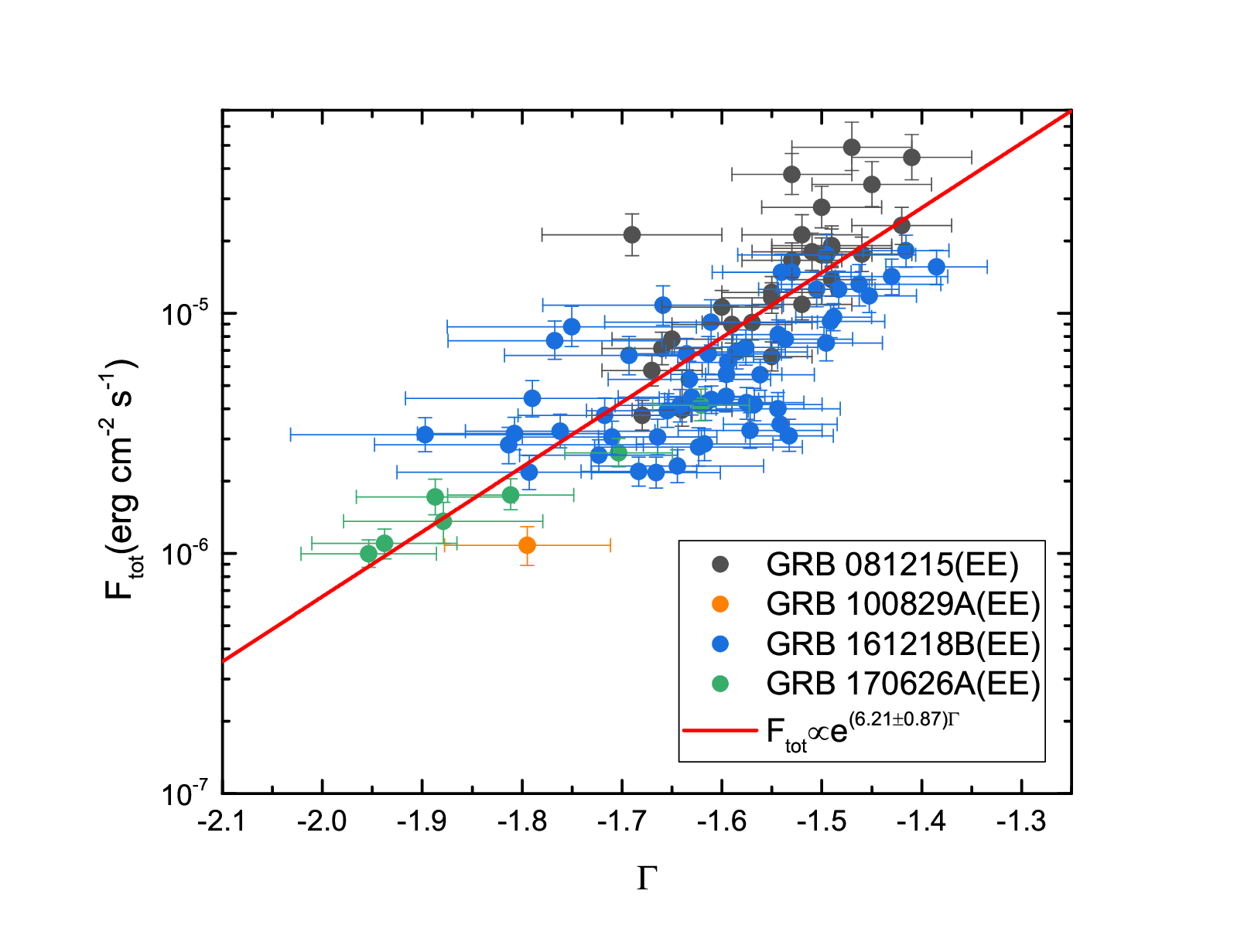}
 \caption{$F_{\rm tot}-\Gamma$ correlations of both ME and EE phases for four GRBs (081215, 100829A, 161218B, and 170626A) of our sample. The solid red lines are the best fit with exponential function.}
\label{fig:GRB F-GAMA}
\end{figure}

Figure \ref{fig:GRB F-GAMA} shows the correlation between $F_{\rm tot}$ and $\Gamma$ for both ME and EE phases of those four GRBs, and we employ the function proposed by \citet{2019MNRAS.484.1912R} to fit the correlations. For the ME phase, one has
\begin{eqnarray}
F_{\rm tot}=(0.42 \pm 0.35) \mathrm{exp}(6.44 \pm 0.62){\Gamma}
\end{eqnarray}
with Spearman correlation coefficient $r=0.83$ and $p < 0.0001$. It suggests that a harder power-law index tends to have a larger total flux. This is consistent with the discovery of the $F_{\rm tot}-\alpha$ relation in \citet{2019MNRAS.484.1912R}. For the EE phase, one has
\begin{eqnarray}
F_{\rm tot}=(0.16 \pm 0.2) \mathrm{exp}(6.21 \pm 0.87){\Gamma}
\end{eqnarray}
with $r=0.79$ and $p < 0.0001$. It is clear that the exponential index of ME and EE are quite close for those two correlations, which suggests that the spectral evolution of both ME and EE seems to share similar behavior.

\subsubsection{$F_{\rm tot}-kT$}
Another well-studied and strong correlation is the $F_{\rm tot}$ and $E_{\rm p}$ obtained from CPL or Band models \citep{1983Natur.306..451G,1994ApJ...422..260K,2001ApJ...548..770B,2012ApJ...756..112L}. In fact, the $E_{\rm p}$ obtained from CPL or Band models seems to correspond to $kT$ obtained from the thermal emission. So, we further investigate the relations between $F_{\rm tot}$ and $kT$ for the four GRBs.

\begin{figure}
\centering
 \includegraphics [angle=0,scale=0.3] {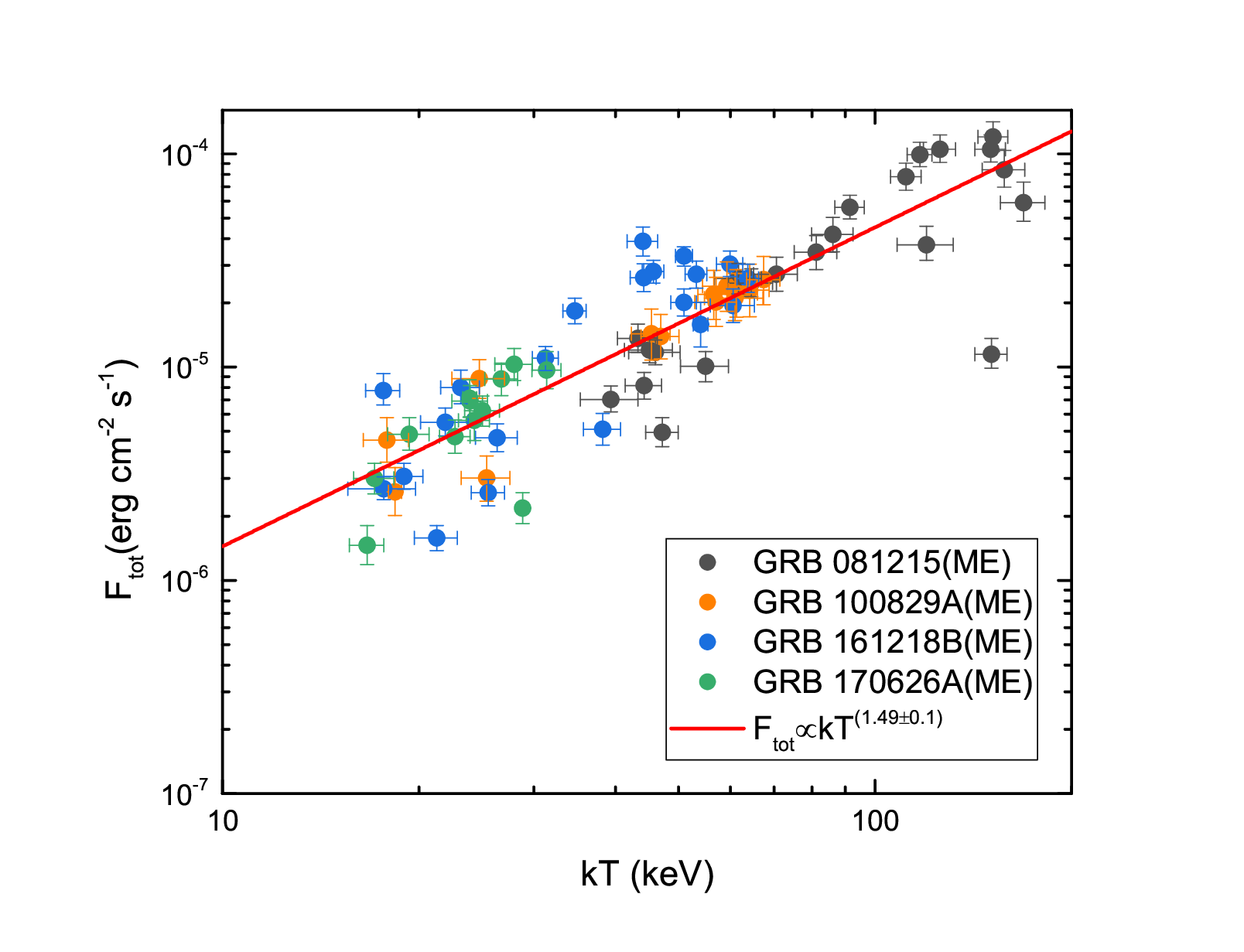}
\includegraphics [angle=0,scale=0.3] {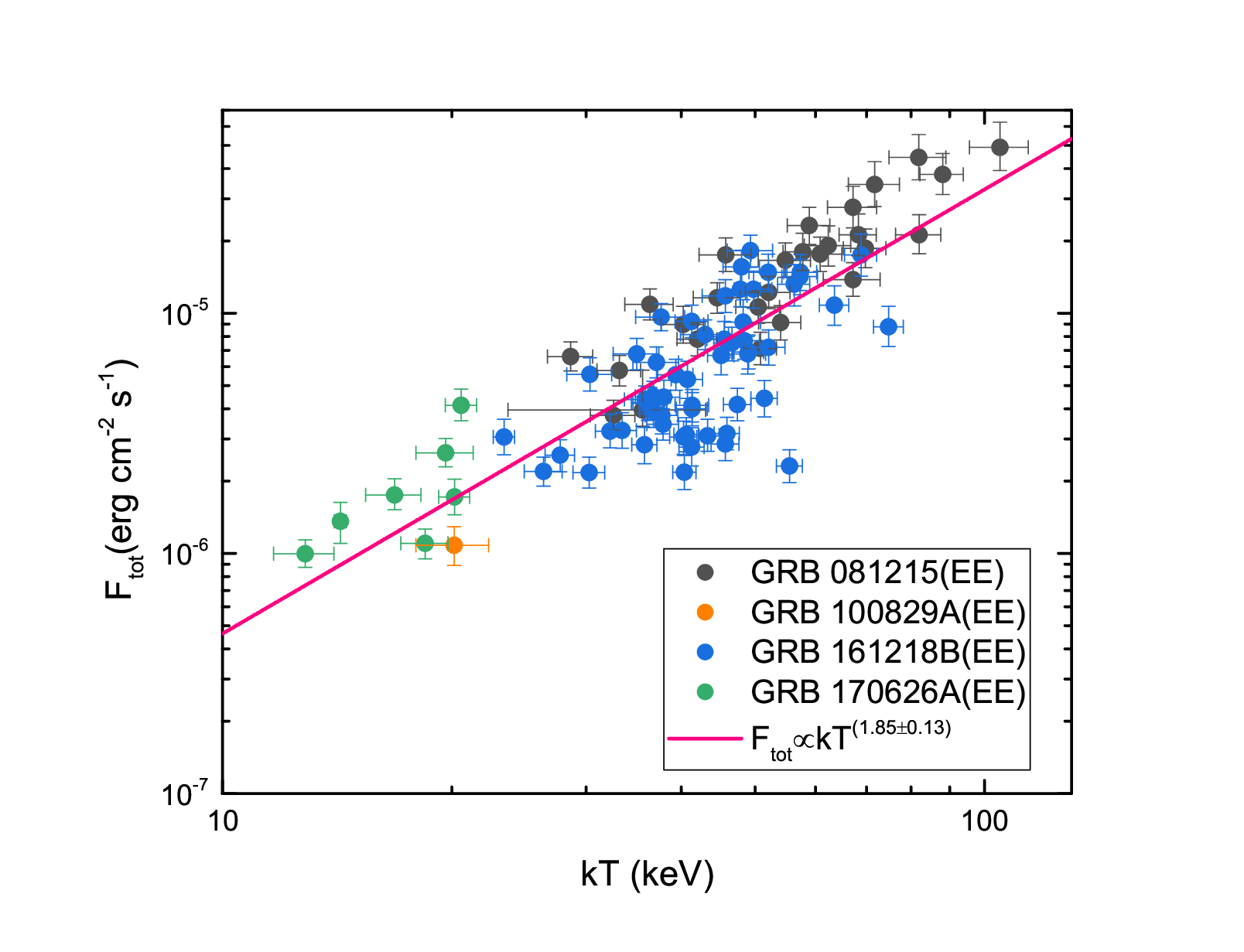}
 \caption{$F_{\rm tot}-kT$ correlations of both ME and EE phases for four GRBs (081215, 100829A, 161218B, and 170626A) of our sample. The solid red lines are the best fit with power-law function.}
\label{fig:GRB F-KT}
\end{figure}

Figure \ref{fig:GRB F-KT} presents a correlation between $F_{\rm tot}$ and $kT$ for both ME and EE phases. For the ME phase, one has
\begin{eqnarray}
\log F_{\rm tot}=(1.49\pm 0.1)\log~(kT)-(7.33\pm 0.16)
\end{eqnarray}
with $r=0.87$ and $p < 0.0001$. It suggests that a higher temperature tends to have a larger total flux. Moreover, for the EE phase, one has
\begin{eqnarray}
\log F_{\rm tot}=(1.85\pm 0.13)\log~(kT)-(8.18\pm 0.21)
\end{eqnarray}
with $r=0.81$ and $p < 0.0001$. It is found that there is a small difference, but not a large one, between those two correlations for ME and EE. This difference may be caused by the occurrence of a break phenomenon in the region above 100 KeV in ME, or from the selection effect due to limited GRB samples. It suggests that both ME and EE phases do not seem to be significantly different statistically.

\section{Conclusion and discussion}{\label{s4}}
In recent years, several peculiar and nearby long-duration GRBs, which are identified as SGRBs with extended emission, have attracted great attention, such as GRB 211227A \citep{2022ApJ...931L..23L}, GRB 211211A \citep{2022Natur.612..223R,2022Natur.612..232Y,2022Natur.612..228T,2023ApJ...943..146C,2023NatAs...7...67G}, and GRB 230307A \citep{2023arXiv230705689S,2023arXiv231007205Y,2024MNRAS.529L..67D,2024ApJ...962L..27D}. Both nonthermal and thermal emissions in GRB 211211A are found to exist, and this suggests that a potential hybrid jet should exist in the prompt emission \citep{2023ApJ...943..146C,2024ApJ...969...26P}.

In this paper, by adopting the criterion of GRB 211211A, we collect the samples of 36 GRBs observed by Fermi/GBM during 15 yr of operation are similar to GRB 211211A, and identify them in the category of SGRBs with extended emission. Then, we adopt the Bayesian method to present a comprehensive spectral analysis for 36 GRBs in our sample and try to identify their jet composition. The time-integrated spectra of 36 GRBs in our sample can be fitted by a nonthermal component (e.g., CPL or Band model), but 20 out of 36 GRBs, the low-energy index $\alpha$ of spectra exceeds the death line of synchrotron emission ($\alpha$=-2/3, \citealt{1998ApJ...506L..23P}). It is very difficult to explain by invoking synchrotron emission, and suggests that a significant contribution from the thermal emission of the fireball photosphere should be considered \citep{2010ApJ...709L.172R,2000ApJ...530..292M}.

Our aim is to consider the thermal component in the prompt emission of those 20 GRBs with $\alpha$ exceeding the death line of synchrotron emission  rather than using the CPL or Band models, and try to solve the problem of death line in nonthermal synchrotron emission. Our results are summarized as follows:

(1) There are 20 out of 36 GRBs ($\sim$55.6\%) for which $\alpha$ values exceed the death line of synchrotron emission within either ME or EE phases. Considering the thermal component in the prompt emission of those 20 GRBs can soften the $\alpha$ values of those 20 GRBs, and suggests that the thermal component may exist in the prompt emission.

(2) We focus on analyzing four bright GRBs with time-integrated spectra for which $\alpha$ exceeds the death line of synchrotron emission and find that the PL$+$BB model consistently provided the best fit in their time-resolved spectra instead of CPL or Band models. This indicates the persistent presence of a quasi-thermal component.

(3) For those four bright GRBs, the evolution of thermal component parameters ($kT$, $F_{\rm BB}$) exhibits certain trends (such as 'tracking' the light curve or 'hard to soft'). $F_{\rm BB}/F_{\rm tot}$ typically exhibits an evolutionary trend and seems to be track the pulses of the light curve, but not significantly.

(4) Two strong positive correlations do exist in the time-resolved spectra of ME and EE phases in the four GRBs, e.g., $F_{\rm tot}-\Gamma$ and $F_{\rm tot}-kT$. Moreover, the close values of indices between the ME and EE phase, together with the high correlation coefficients, suggests that the spectral evolution of both ME and EE share similar behaviors, and there seems to be no significant difference, possibly from the same physical origin.

On the other hand, the hard low-energy spectral index $\alpha$ is used as an argument for a photospheric origin of GRB spectra. Previous studies have also suggested that some LGRBs have thermal emission. For example, \citet{2019MNRAS.487.5508A} found that more than a quarter of the bursts may exist the photosphere emission within nondissipative outflow. However, the thermal component should not look like a simple Planck spectra, but rather have a more complicated shape \citep{2020ApJ...893..128A}.

From the physical point of view, a hybrid jet in the GRB outflow should exist \citep{2015ApJ...801...2}, e.g., a quasi-thermal comes from photosphere emission in fireball model \citep{1986ApJ...308L..43P,1986ApJ...308L..47G,2010ApJ...709L.172R,2012MNRAS.420..468P}, and nonthermal Poynting flux comes from magnetic dissipation in the ICMART model \citep{2011ApJ...726...90Z}. The discovery of thermal emission in the four bright GRBs should be attributed to photosphere emission, and the PL component has been extensively studied within the context of both internal \citep{2009ApJ...705L.191A,2009A&A...498..677B,2010A&A...524A..92C,2011ApJ...739..103A,2020ApJ...891..106A} and external \citep{1994MNRAS.269L..41M,2009MNRAS.400L..75K,2014ApJ...788...36B,2017ApJ...848...94F} dissipation models. However, the external model may encounter difficulty in explaining the correlated temporal behavior observed between GeV and keV-MeV emissions in certain GRBs \citep{2017ApJ...844...56T}. Within the framework of internal dissipation model, the PL component may be from the synchrotron radiation \citep{2010PThPh.124..667I}, Comptonized thermal, synchrotron self-Compton \citep{1994ApJ...430L..93R,2007ApJ...655..762A}, photopion production and Bethe-Heitler pair production of relativistic protons \citep{2018ApJ...857...24W}, or the inverse Compton scattering of high-energy electrons \citep{2018ApJ...857...24W,2021ApJ...922..255T}.

From a theoretical point of view, \citet{2017ApJ...849...47L} proposed two jet emission mechanisms, namely the $\nu \bar{\nu}$ annihilation and Blandford-Znajek (BZ, \citealt{1977MNRAS.179..433B}) processes, which are considered to provide power to the thermal and nonthermal components of relativistic jets, respectively. However, the exact workings of these two mechanisms are still unclear. In any case, understanding the physical process based solely on the properties of spectral is, of course, not an easy task. We therefore expect to follow-up multiband observations for those short GRB with EE events in the future.

\begin{acknowledgements}
We are very grateful to the referee for the careful and thoughtful suggestions that have helped improve this manuscript substantially. We acknowledge the use of the public data from the Fermi/GBM data archive. This work is supported by the Guangxi Science Foundation (grant No. 2017GXNSFFA198008), the National Natural Science Foundation of China (grant Nos. 11922301, and 12133003), the Program of Bagui Scholars Program (LHJ), and the Guangxi Talent Program (“Highland of Innovation Talents”).

\end{acknowledgements}


\clearpage
\startlongtable
\begin{deluxetable*}{ccccccccccc}
\tablewidth{0pt}
\tabletypesize{\footnotesize}
\tablecaption{The fitting results of the time-integrated spectra with Band and CPL models for our sample.}
\tablehead{
GRB
&Detector
&Phase
&$t_1\sim t_2$
&$\quad$
&CPL
&$\quad$
&Band
&$\quad$
&$\Delta$BIC
&Preferred\\
$\quad$
&$\quad$
&$\quad$
&$\quad$
&$\alpha$
&$E_c$(keV)
&$\alpha$
&$E_p$(keV)
&$\beta$
&$\quad$
&Model
}

\startdata
080807 	&	n1,n2,n5,b0	&	ME	&	0.0 	$\sim$	2.0 	&$	-0.60 	_{	-0.11 	}^{+	0.11 	}$&$	467.97 	_{	-125.43 	}^{+	119.80 	}$&$	-0.53 	_{	-0.14 	}^{+	0.14 	}$&$	549.28 	_{	-129.36 	}^{+	130.95 	}$&$	-2.39 	_{	-0.39 	}^{+	0.38 	}$&	6.47 	&	CPL	\\
$\quad$	&	$\quad$	&	EE	&	2.0 	$\sim$	20.0 	&$	-0.77 	_{	-0.17 	}^{+	0.17 	}$&$	304.71 	_{	-118.41 	}^{+	108.35 	}$&$	-0.63 	_{	-0.32 	}^{+	0.38 	}$&$	302.81 	_{	-146.79 	}^{+	141.10 	}$&$	-2.27 	_{	-0.52 	}^{+	0.42 	}$&	-0.26 	&	Band	\\
$\quad$	&	$\quad$	&	Tot	&	0.0 	$\sim$	20.0 	&$	-0.77 	_{	-0.12 	}^{+	0.12 	}$&$	393.08 	_{	-119.73 	}^{+	116.46 	}$&$	-0.69 	_{	-0.20 	}^{+	0.18 	}$&$	411.38 	_{	-146.22 	}^{+	137.85 	}$&$	-2.25 	_{	-0.48 	}^{+	0.40 	}$&	1.27 	&	CPL	\\
081110 	&	n7,b8,b1	&	ME	&	-1.0 	$\sim$	9.0 	&$	-0.91 	_{	-0.06 	}^{+	0.06 	}$&$	411.57 	_{	-67.27 	}^{+	66.74 	}$&$	-0.84 	_{	-0.09 	}^{+	0.08 	}$&$	373.75 	_{	-73.57 	}^{+	72.01 	}$&$	-2.57 	_{	-0.54 	}^{+	0.51 	}$&	4.26 	&	CPL	\\
$\quad$	&	$\quad$	&	EE	&	9.0 	$\sim$	19.0 	&$	-0.91 	_{	-0.28 	}^{+	0.28 	}$&$	136.25 	_{	-70.60 	}^{+	48.55 	}$&	——	&	——	&	——	&	——	&	CPL	\\
$\quad$	&	$\quad$	&	Tot	&	-1.0 	$\sim$	19.0 	&$	-1.00 	_{	-0.06 	}^{+	0.06 	}$&$	402.25 	_{	-74.60 	}^{+	72.96 	}$&$	-0.91 	_{	-0.10 	}^{+	0.10 	}$&$	307.92 	_{	-76.18 	}^{+	74.09 	}$&$	-2.37 	_{	-0.44 	}^{+	0.42 	}$&	1.79 	&	CPL	\\
081129 	&	na,nb,b1	&	ME	&	-2.0 	$\sim$	5.0 	&$	-0.88 	_{	-0.06 	}^{+	0.06 	}$&$	291.05 	_{	-44.04 	}^{+	43.42 	}$&$	-0.81 	_{	-0.08 	}^{+	0.08 	}$&$	276.66 	_{	-37.93 	}^{+	38.38 	}$&$	-2.55 	_{	-0.33 	}^{+	0.33 	}$&	2.07 	&	CPL	\\
$\quad$	&	$\quad$	&	EE	&	5.0 	$\sim$	30.0 	&$	-1.02 	_{	-0.07 	}^{+	0.07 	}$&$	278.63 	_{	-50.80 	}^{+	50.67 	}$&$	-1.00 	_{	-0.08 	}^{+	0.08 	}$&$	247.94 	_{	-35.25 	}^{+	35.13 	}$&$	-2.65 	_{	-0.43 	}^{+	0.42 	}$&	3.67 	&	CPL	\\
\enddata
\label{Tablel}
\end{deluxetable*}

\begin{table*}
\centering
\renewcommand\tabcolsep{2.2pt}
\renewcommand\arraystretch{1}
\caption{Comparisons of $\Delta$BIC for different models of time-resolved spectra for GRB 081215.}
\begin{tabular}{cccccccc} 
\hline\hline
$t_{1}\sim t_{2}$&$\alpha$&$\quad$&$\quad$&$\Delta$BIC&$\quad$&$\quad$&\rm Best Model \\
(s)&$\quad$&Band&BB&PL+BB&CPL+BB&Band+BB& \\

\hline
0.80 	$\sim$	1.25 	&$	-0.64 	_{	-0.06 	}^{+	0.06 	}$&	9.20 	&	165.56 	&	-40.81 	&	Non-Con.	&	Non-Con.	&	PL+BB	\\
1.25 	$\sim$	1.33 	&$	-0.75 	_{	-0.06 	}^{+	0.06 	}$&	-16.37 	&	138.62 	&	-71.71 	&	Non-Con.	&	Non-Con.	&	PL+BB	\\
1.33 	$\sim$	1.38 	&$	-0.60 	_{	-0.06 	}^{+	0.06 	}$&	-2.04 	&	90.22 	&	-42.77 	&	Non-Con.	&	Non-Con.	&	PL+BB	\\
1.38 	$\sim$	1.41 	&$	-0.57 	_{	-0.06 	}^{+	0.06 	}$&	-6.67 	&	110.90 	&	-58.55 	&	Non-Con.	&	Non-Con.	&	PL+BB	\\
1.41 	$\sim$	1.45 	&$	-0.55 	_{	-0.05 	}^{+	0.04 	}$&	0.27 	&	245.21 	&	-15.65 	&	Non-Con.	&	Non-Con.	&	PL+BB	\\
1.45 	$\sim$	1.48 	&$	-0.47 	_{	-0.06 	}^{+	0.06 	}$&	-13.55 	&	160.54 	&	-44.60 	&	Non-Con.	&	Non-Con.	&	PL+BB	\\
\hline
\end{tabular}
\label{Table2}
\end{table*}

\begin{table*}
\centering
\renewcommand\tabcolsep{2.2pt}
\renewcommand\arraystretch{1}
\caption{The fitting results of time-resolved spectra with CPL and PL$+$BB models in GRB 081215 prompt emission.}
\begin{tabular}{ccccccccccc}
\hline\hline
$t_{1}\sim t_{2}$
&$\quad$
&CPL
&BIC
&RSS/dof
&$\quad$
&PL+BB
&BIC
&RSS/dof
&$F_{BB}$
&$F_{BB}/F_{tot}$
\\
$\quad$
&$\alpha$
&$E_{c}$(keV)
&
&
&$\Gamma$
&kT(keV)
&
&
&($\rm 10^{-6} cm^{-2} s^{-1}$)
&\\
\hline
0.80 	$\sim$	1.25 	&$	-0.64 	_{	-0.06 	}^{+	0.06 	}$&$	924.16 	_{	-138.74 	}^{+	138.28 	}$&	1465.44 	&	411/466	&$	-1.45 	_{	-0.04 	}^{+	0.04 	}$&$	150.68 	_{	-8.52 	}^{+	8.45 	}$&	1424.63 	&	439/465	&$	5.80 	^{+	1.93 	}_{	-1.48 	}$&$	0.50 	^{+	0.19 	}_{-	0.15 	}$	\\
1.25 	$\sim$	1.33 	&$	-0.75 	_{	-0.06 	}^{+	0.06 	}$&$	1301.48 	_{	-249.91 	}^{+	249.00 	}$&	303.69 	&	331/466	&$	-1.33 	_{	-0.03 	}^{+	0.03 	}$&$	119.92 	_{	-11.82 	}^{+	11.89 	}$&	231.98 	&	331/465	&$	11.30 	^{+	7.53 	}_{	-4.43 	}$&$	0.30 	^{+	0.21 	}_{-	0.13 	}$	\\
1.33 	$\sim$	1.38 	&$	-0.60 	_{	-0.06 	}^{+	0.06 	}$&$	1288.82 	_{	-186.75 	}^{+	187.84 	}$&	240.44 	&	305/466	&$	-1.30 	_{	-0.04 	}^{+	0.04 	}$&$	168.76 	_{	-13.25 	}^{+	13.28 	}$&	197.67 	&	328/465	&$	27.90 	^{+	13.50 	}_{	-9.15 	}$&$	0.47 	^{+	0.26 	}_{-	0.18 	}$	\\
1.38 	$\sim$	1.41 	&$	-0.57 	_{	-0.06 	}^{+	0.06 	}$&$	1090.53 	_{	-177.05 	}^{+	177.67 	}$&	144.54 	&	322/466	&$	-1.32 	_{	-0.04 	}^{+	0.04 	}$&$	157.66 	_{	-11.83 	}^{+	11.83 	}$&	85.99 	&	331/465	&$	39.00 	^{+	18.20 	}_{	-12.10 	}$&$	0.46 	^{+	0.24 	}_{-	0.16 	}$	\\
1.41 	$\sim$	1.45 	&$	-0.55 	_{	-0.05 	}^{+	0.04 	}$&$	1029.84 	_{	-107.55 	}^{+	108.01 	}$&	183.44 	&	334/466	&$	-1.33 	_{	-0.03 	}^{+	0.03 	}$&$	150.28 	_{	-8.08 	}^{+	8.19 	}$&	167.79 	&	368/465	&$	52.10 	^{+	16.10 	}_{	-12.10 	}$&$	0.49 	^{+	0.17 	}_{-	0.13 	}$	\\
1.45 	$\sim$	1.48 	&$	-0.47 	_{	-0.06 	}^{+	0.06 	}$&$	772.70 	_{	-104.58 	}^{+	103.98 	}$&	176.91 	&	343/466	&$	-1.34 	_{	-0.03 	}^{+	0.04 	}$&$	151.69 	_{	-7.93 	}^{+	8.03 	}$&	132.31 	&	339/465	&$	63.40 	^{+	19.90 	}_{	-14.80 	}$&$	0.53 	^{+	0.19 	}_{-	0.14 	}$	\\
\hline
\end{tabular}
\label{Table3}
\end{table*}

\clearpage
\begin{table*}
\centering
\renewcommand\tabcolsep{2.2pt}
\renewcommand\arraystretch{1}
\caption{Comparisons of $\Delta$BIC for different models of time-resolved spectra for GRB 100829A.}
\begin{tabular}{cccccccc} 
\hline\hline
$t_{1}\sim t_{2}$&$\alpha$&$\quad$&$\quad$&$\Delta$BIC&$\quad$&$\quad$&\rm Best Model \\
(s)&$\quad$&Band&BB&PL+BB&CPL+BB&Band+BB& \\

\hline
-0.60 	$\sim$	0.28 	&$	-0.14 	_{	-0.19 	}^{+	0.19 	}$&	9.43 	&	13.87 	&	-23.04 	&	Non-Con.	&	Non-Con.	&	PL+BB	\\
0.28 	$\sim$	0.55 	&$	-0.34 	_{	-0.22 	}^{+	0.22 	}$&	6.87 	&	11.86 	&	-18.60 	&	Non-Con.	&	Non-Con.	&	PL+BB	\\
0.55 	$\sim$	0.63 	&$	-0.15 	_{	-0.18 	}^{+	0.18 	}$&	7.15 	&	0.97 	&	-33.66 	&	Non-Con.	&	Non-Con.	&	PL+BB	\\
0.63 	$\sim$	0.68 	&$	-0.36 	_{	-0.14 	}^{+	0.14 	}$&	13.03 	&	32.99 	&	-9.73 	&	Non-Con.	&	Non-Con.	&	PL+BB	\\
0.68 	$\sim$	0.73 	&$	-0.23 	_{	-0.14 	}^{+	0.14 	}$&	12.97 	&	17.02 	&	-13.62 	&	Non-Con.	&	Non-Con.	&	PL+BB	\\
0.73 	$\sim$	0.78 	&$	-0.36 	_{	-0.13 	}^{+	0.13 	}$&	10.33 	&	30.81 	&	-2.79 	&	Non-Con.	&	Non-Con.	&	PL+BB(CPL)	\\
0.78 	$\sim$	0.86 	&$	-0.55 	_{	-0.12 	}^{+	0.12 	}$&	9.67 	&	57.88 	&	-6.04 	&	Non-Con.	&	Non-Con.	&	PL+BB	\\
0.86 	$\sim$	0.91 	&$	-0.33 	_{	-0.13 	}^{+	0.13 	}$&	14.13 	&	42.32 	&	0.26 	&	Non-Con.	&	Non-Con.	&	CPL(PL+BB)	\\
0.91 	$\sim$	0.99 	&$	-0.28 	_{	-0.16 	}^{+	0.16 	}$&	7.74 	&	20.51 	&	-15.52 	&	Non-Con.	&	Non-Con.	&	PL+BB	\\
0.99 	$\sim$	1.17 	&$	-0.49 	_{	-0.17 	}^{+	0.17 	}$&	-1.23 	&	30.84 	&	-16.00 	&	Non-Con.	&	Non-Con.	&	PL+BB	\\
1.17 	$\sim$	1.59 	&$	-0.54 	_{	-0.16 	}^{+	0.16 	}$&	9.30 	&	38.32 	&	-8.20 	&	Non-Con.	&	Non-Con.	&	PL+BB	\\
1.59 	$\sim$	1.63 	&$	0.00 	_{	-0.17 	}^{+	0.17 	}$&	10.42 	&	-10.61 	&	-25.37 	&	Non-Con.	&	Non-Con.	&	PL+BB	\\
1.63 	$\sim$	1.67 	&$	0.18 	_{	-0.19 	}^{+	0.19 	}$&	14.61 	&	-18.30 	&	-31.49 	&	Non-Con.	&	Non-Con.	&	PL+BB	\\
1.67 	$\sim$	6.78 	&$	-0.98 	_{	-0.14 	}^{+	0.14 	}$&	4.96	&	103.69	&	-28.51	&	Non-Con.	&	Non-Con.	&	PL+BB	\\

\hline
\end{tabular}
\label{Table4}
\end{table*}


\newpage
\begin{table*}
\centering
\renewcommand\tabcolsep{2.2pt}
\renewcommand\arraystretch{1}
\caption{The fitting results of time-resolved spectra with CPL and PL$+$BB models in GRB 100829A prompt emission.}
\begin{tabular}{ccccccccccc}
\hline\hline
$t_{1}\sim t_{2}$
&$\quad$
&CPL
&BIC
&RSS/dof
&$\quad$
&PL+BB
&BIC
&RSS/dof
&$F_{BB}$
&$F_{BB}/F_{tot}$
\\
$\quad$
&$\alpha$
&$E_{c}$(keV)
&
&
&$\Gamma$
&kT(keV)
&
&
&($\rm 10^{-6} cm^{-2} s^{-1}$)
&
\\
\hline
-0.60 	$\sim$	0.28 	&$	-0.14 	_{	-0.19 	}^{+	0.19 	}$&$	48.27 	_{	-7.70 	}^{+	7.59 	}$&	1009.63 	&	249/237	&$	-1.66 	_{	-0.09 	}^{+	0.09 	}$&$	18.38 	_{	-0.43 	}^{+	2.28 	}$&	986.60 	&	245/236	&$	0.64 	^{+	0.29 	}_{	-0.30 	}$&$	0.25 	^{+	0.13 	}_{-	0.13 	}$	\\
0.28 	$\sim$	0.55 	&$	-0.34 	_{	-0.22 	}^{+	0.22 	}$&$	53.53 	_{	-11.23 	}^{+	11.18 	}$&	328.02 	&	180/237	&$	-1.72 	_{	-0.10 	}^{+	0.10 	}$&$	17.87 	_{	-1.42 	}^{+	1.41 	}$&	309.43 	&	170/236	&$	1.19 	^{+	0.66 	}_{	-0.45 	}$&$	0.26 	^{+	0.11 	}_{-	0.16 	}$	\\
0.55 	$\sim$	0.63 	&$	-0.15 	_{	-0.18 	}^{+	0.18 	}$&$	123.73 	_{	-22.71 	}^{+	22.64 	}$&	44.00 	&	184/237	&$	-1.65 	_{	-0.13 	}^{+	0.15 	}$&$	45.41 	_{	-3.10 	}^{+	3.08 	}$&	10.34 	&	169/236	&$	6.84 	^{+	2.95 	}_{	-2.15 	}$&$	0.48 	^{+	0.19 	}_{-	0.25 	}$	\\
0.63 	$\sim$	0.68 	&$	-0.36 	_{	-0.14 	}^{+	0.14 	}$&$	184.72 	_{	-33.16 	}^{+	32.69 	}$&	-74.93 	&	128/237	&$	-1.71 	_{	-0.11 	}^{+	0.11 	}$&$	57.07 	_{	-4.15 	}^{+	4.09 	}$&	-84.66 	&	143/236	&$	11.20 	^{+	5.52 	}_{	-3.58 	}$&$	0.55 	^{+	0.22 	}_{-	0.32 	}$	\\
0.68 	$\sim$	0.73 	&$	-0.23 	_{	-0.14 	}^{+	0.14 	}$&$	194.23 	_{	-33.17 	}^{+	32.80 	}$&	-104.33 	&	117/237	&$	-1.68 	_{	-0.13 	}^{+	0.14 	}$&$	64.23 	_{	-4.53 	}^{+	4.53 	}$&	-117.95 	&	132/236	&$	14.50 	^{+	6.72 	}_{	-4.58 	}$&$	0.63 	^{+	0.25 	}_{-	0.36 	}$	\\
0.73 	$\sim$	0.78 	&$	-0.36 	_{	-0.13 	}^{+	0.13 	}$&$	223.54 	_{	-40.27 	}^{+	40.37 	}$&	-84.80 	&	153/237	&$	-1.55 	_{	-0.08 	}^{+	0.09 	}$&$	59.25 	_{	-4.89 	}^{+	4.86 	}$&	-87.59 	&	177/236	&$	11.40 	^{+	6.03 	}_{	-4.10 	}$&$	0.48 	^{+	0.21 	}_{-	0.29 	}$	\\
\hline
\end{tabular}
\label{Table5}
\end{table*}

\clearpage
\startlongtable
\begin{deluxetable*}{cccccccc}
\tablewidth{0pt}
\tabletypesize{\footnotesize}
\tablecaption{Comparisons of $\Delta$BIC for different models of time-resolved spectra for GRB 161218B.}
\tablehead{
$t_{1}\sim t_{2}$
&$\alpha$
&$\quad$
&$\quad$
&$\Delta$BIC
&$\quad$
&$\quad$
&\rm Best Model
\\
(s)
&
&Band
&BB
&PL+BB
&CPL+BB
&Band+BB
&
}
\startdata
0.00 	$\sim$	0.71 	&$	-1.16 	_{	-0.10 	}^{+	0.10 	}$&	-3.72 	&	207.00 	&	-34.93 	&	Non-Con.	&	Non-Con.	&	PL+BB	\\
0.71 	$\sim$	1.00 	&$	-0.71 	_{	-0.11 	}^{+	0.11 	}$&	9.96 	&	126.35 	&	-41.92 	&	Non-Con.	&	Non-Con.	&	PL+BB	\\
1.00 	$\sim$	1.47 	&$	-0.67 	_{	-0.11 	}^{+	0.11 	}$&	24.64 	&	141.83 	&	-21.14 	&	Non-Con.	&	Non-Con.	&	PL+BB	\\
1.47 	$\sim$	1.69 	&$	-0.66 	_{	-0.10 	}^{+	0.10 	}$&	16.62 	&	123.99 	&	-38.96 	&	Non-Con.	&	Non-Con.	&	PL+BB	\\
1.69 	$\sim$	1.77 	&$	-0.52 	_{	-0.09 	}^{+	0.09 	}$&	15.11 	&	93.31 	&	-33.16 	&	Non-Con.	&	Non-Con.	&	PL+BB	\\
1.77 	$\sim$	1.84 	&$	-0.46 	_{	-0.10 	}^{+	0.10 	}$&	18.68 	&	75.55 	&	-54.81 	&	Non-Con.	&	Non-Con.	&	PL+BB	\\
1.84 	$\sim$	1.92 	&$	-0.27 	_{	-0.09 	}^{+	0.09 	}$&	19.94 	&	104.05 	&	-39.57 	&	Non-Con.	&	Non-Con.	&	PL+BB	\\
1.92 	$\sim$	1.99 	&$	-0.32 	_{	-0.08 	}^{+	0.08 	}$&	17.16 	&	132.05 	&	-22.27 	&	Non-Con.	&	Non-Con.	&	PL+BB	\\
\enddata
\end{deluxetable*}\label{Table6}

\clearpage
\startlongtable
\begin{deluxetable*}{ccccccccccc}
\tablewidth{0pt}
\tabletypesize{\footnotesize}
\tablecaption{The fitting results of time-resolved spectra with CPL and PL$+$BB in GRB 161218B prompt emission.}
\tablehead{
$t_{1}\sim t_{2}$
&$\quad$
&CPL
&BIC
&RSS/dof
&$\quad$
&PL+BB
&BIC
&RSS/dof
&$F_{BB}$
&$F_{BB}/F_{tot}$
\\
$\quad$
&$\alpha$
&$E_{c}$(keV)
&
&
&$\Gamma$
&kT(keV)
&
&
&($\rm 10^{-6} cm^{-2} s^{-1}$)
&
}
\startdata
0.00 	$\sim$	0.71 	&$	-1.16 	_{	-0.10 	}^{+	0.10 	}$&$	131.92 	_{	-27.04 	}^{+	26.62 	}$&	1806.55 	&	372 	/	470	&$	-1.81 	_{	-0.05 	}^{+	0.05 	}$&$	17.66 	_{	-2.09 	}^{+	2.10 	}$&	1771.62 	&	371 	/	469	&$	0.30 	^{+	0.29 	}_{	-0.15 	}$&$	0.11 	^{+	0.11 	}_{-	0.06 	}$	\\
0.71 	$\sim$	1.00 	&$	-0.71 	_{	-0.11 	}^{+	0.11 	}$&$	103.17 	_{	-16.36 	}^{+	16.44 	}$&	887.45 	&	365 	/	470	&$	-1.74 	_{	-0.06 	}^{+	0.06 	}$&$	26.35 	_{	-1.92 	}^{+	1.91 	}$&	845.53 	&	361 	/	469	&$	0.98 	^{+	0.46 	}_{	-0.32 	}$&$	0.21 	^{+	0.10 	}_{-	0.08 	}$	\\
1.00 	$\sim$	1.47 	&$	-0.67 	_{	-0.11 	}^{+	0.11 	}$&$	84.25 	_{	-11.02 	}^{+	11.07 	}$&	1302.45 	&	336 	/	470	&$	-1.92 	_{	-0.07 	}^{+	0.08 	}$&$	25.54 	_{	-1.50 	}^{+	1.50 	}$&	1281.31 	&	361 	/	469	&$	0.80 	^{+	0.30 	}_{	-0.23 	}$&$	0.31 	^{+	0.13 	}_{-	0.10 	}$	\\
1.47 	$\sim$	1.69 	&$	-0.66 	_{	-0.10 	}^{+	0.10 	}$&$	151.79 	_{	-22.70 	}^{+	22.35 	}$&	774.91 	&	321 	/	470	&$	-1.78 	_{	-0.07 	}^{+	0.07 	}$&$	38.24 	_{	-2.51 	}^{+	2.50 	}$&	735.95 	&	343 	/	469	&$	1.81 	^{+	0.71 	}_{	-0.53 	}$&$	0.35 	^{+	0.15 	}_{-	0.12 	}$	\\
1.69 	$\sim$	1.77 	&$	-0.52 	_{	-0.09 	}^{+	0.09 	}$&$	261.00 	_{	-40.37 	}^{+	39.88 	}$&	154.37 	&	314 	/	470	&$	-1.51 	_{	-0.06 	}^{+	0.05 	}$&$	60.63 	_{	-4.70 	}^{+	4.74 	}$&	121.21 	&	330 	/	469	&$	6.28 	^{+	3.11 	}_{	-2.08 	}$&$	0.32 	^{+	0.17 	}_{-	0.12 	}$	\\
1.77 	$\sim$	1.84 	&$	-0.46 	_{	-0.10 	}^{+	0.10 	}$&$	202.92 	_{	-28.74 	}^{+	28.61 	}$&	158.89 	&	300 	/	470	&$	-1.57 	_{	-0.07 	}^{+	0.07 	}$&$	54.03 	_{	-1.39 	}^{+	6.69 	}$&	104.07 	&	302 	/	469	&$	5.62 	^{+	2.36 	}_{	-2.31 	}$&$	0.35 	^{+	0.18 	}_{-	0.16 	}$	\\
1.84 	$\sim$	1.92 	&$	-0.27 	_{	-0.09 	}^{+	0.09 	}$&$	152.42 	_{	-16.26 	}^{+	16.33 	}$&	212.06 	&	312 	/	470	&$	-1.59 	_{	-0.06 	}^{+	0.06 	}$&$	50.93 	_{	-2.28 	}^{+	2.30 	}$&	172.49 	&	327 	/	469	&$	8.80 	^{+	2.41 	}_{	-1.87 	}$&$	0.44 	^{+	0.14 	}_{-	0.11 	}$	\\
\enddata
\end{deluxetable*}\label{Table7}

\clearpage
\begin{table*}
\centering
\renewcommand\tabcolsep{2.2pt}
\renewcommand\arraystretch{1}
\caption{Comparisons of $\Delta$BIC for different models of time-resolved spectra for GRB 170626A.}
\begin{tabular}{cccccccc} 
\hline\hline
$t_{1}\sim t_{2}$&$\alpha$&$\quad$&$\quad$&$\Delta$BIC&$\quad$&$\quad$&\rm Best Model \\
(s)&$\quad$&Band&BB&PL+BB&CPL+BB&Band+BB& \\

\hline
-0.60 	$\sim$	1.17 	&$	-0.45 	_{	-0.09 	}^{+	0.09 	}$&	13.96 	&	Non-Con.	&	-52.66 	&	Non-Con.	&	Non-Con.	&	PL+BB	\\
1.17 	$\sim$	1.26 	&$	-0.39 	_{	-0.13 	}^{+	0.13 	}$&	6.42 	&	37.41 	&	-43.94 	&	Non-Con.	&	Non-Con.	&	PL+BB	\\
1.26 	$\sim$	1.35 	&$	-0.37 	_{	-0.12 	}^{+	0.12 	}$&	18.69 	&	54.69 	&	-29.63 	&	Non-Con.	&	Non-Con.	&	PL+BB	\\
1.35 	$\sim$	1.46 	&$	-0.27 	_{	-0.13 	}^{+	0.13 	}$&	17.42 	&	38.04 	&	-46.62 	&	Non-Con.	&	Non-Con.	&	PL+BB	\\
1.46 	$\sim$	1.55 	&$	-0.06 	_{	-0.16 	}^{+	0.16 	}$&	26.75 	&	-0.17 	&	-41.38 	&	Non-Con.	&	Non-Con.	&	PL+BB	\\
1.55 	$\sim$	1.66 	&$	-0.35 	_{	-0.13 	}^{+	0.13 	}$&	13.14 	&	47.13 	&	-35.85 	&	Non-Con.	&	Non-Con.	&	PL+BB	\\
\hline
\end{tabular}
\label{Table8}
\end{table*}

\begin{table*}
\centering
\renewcommand\tabcolsep{2.2pt}
\renewcommand\arraystretch{1}
\caption{The fitting results of time-resolved spectra with CPL and PL$+$BB in GRB 170626A prompt emission.}
\begin{tabular}{ccccccccccc}
\hline\hline
$t_{1}\sim t_{2}$
&$\quad$
&CPL
&BIC
&RSS/dof
&$\quad$
&PL+BB
&BIC
&RSS/dof
&$F_{BB}$
&$F_{BB}/F_{tot}$
\\
$\quad$
&$\alpha$
&$E_{c}$(keV)
&
&
&$\Gamma$
&kT(keV)
&
&
&($\rm 10^{-6} cm^{-2} s^{-1}$)
&
\\
\hline
-0.60 	$\sim$	1.17 	&$	-0.45 	_{	-0.09 	}^{+	0.09 	}$&$	98.91 	_{	-11.25 	}^{+	11.32 	}$&	3277.99 	&	464/470	&$	-1.60 	_{	-0.05 	}^{+	0.06 	}$&$	28.85 	^{+	2.79 	}_{	-0.39 	}$&	3225.33 	&	466/469	&$	0.46 	^{+	0.15 	}_{	-0.15 	}$&$	0.21 	^{+	0.08 	}_{-	0.07 	}$	\\
1.17 	$\sim$	1.26 	&$	-0.39 	_{	-0.13 	}^{+	0.13 	}$&$	97.89 	_{	-15.11 	}^{+	15.03 	}$&	294.31 	&	327/470	&$	-1.57 	_{	-0.06 	}^{+	0.06 	}$&$	28.00 	^{+	1.83 	}_{	-1.85 	}$&	250.37 	&	324/469	&$	2.46 	^{+	1.06 	}_{	-0.73 	}$&$	0.24 	^{+	0.11 	}_{-	0.08 	}$	\\
1.26 	$\sim$	1.35 	&$	-0.37 	_{	-0.12 	}^{+	0.12 	}$&$	97.45 	_{	-13.35 	}^{+	13.20 	}$&	198.65 	&	285/470	&$	-1.67 	_{	-0.08 	}^{+	0.08 	}$&$	31.40 	^{+	2.67 	}_{	-1.60 	}$&	169.02 	&	301/469	&$	3.10 	^{+	1.36 	}_{	-0.98 	}$&$	0.32 	^{+	0.16 	}_{-	0.12 	}$	\\
1.35 	$\sim$	1.46 	&$	-0.27 	_{	-0.13 	}^{+	0.13 	}$&$	74.88 	_{	-9.45 	}^{+	9.32 	}$&	370.88 	&	342/470	&$	-1.63 	_{	-0.07 	}^{+	0.07 	}$&$	26.77 	^{+	1.53 	}_{	-1.55 	}$&	324.26 	&	328/469	&$	2.49 	^{+	0.90 	}_{	-0.67 	}$&$	0.28 	^{+	0.12 	}_{-	0.09 	}$	\\
1.46 	$\sim$	1.55 	&$	-0.06 	_{	-0.16 	}^{+	0.16 	}$&$	55.45 	_{	-7.01 	}^{+	6.92 	}$&	292.82 	&	303/470	&$	-1.81 	_{	-0.13 	}^{+	0.13 	}$&$	24.32 	^{+	1.31 	}_{	-1.32 	}$&	251.44 	&	312/469	&$	2.58 	^{+	0.87 	}_{	-0.66 	}$&$	0.46 	^{+	0.19 	}_{-	0.15 	}$	\\
1.55 	$\sim$	1.66 	&$	-0.35 	_{	-0.13 	}^{+	0.13 	}$&$	71.78 	_{	-9.36 	}^{+	9.38 	}$&	394.93 	&	347/470	&$	-1.73 	_{	-0.08 	}^{+	0.08 	}$&$	23.93 	^{+	2.57 	}_{	-0.58 	}$&	359.08 	&	348/469	&$	2.16 	^{+	0.89 	}_{	-0.75 	}$&$	0.30 	^{+	0.14 	}_{-	0.12 	}$	\\
\hline
\end{tabular}
\label{Table9}
\end{table*}

\begin{table*}
\centering
\renewcommand\tabcolsep{2.2pt}
\renewcommand\arraystretch{0.9}
\caption{Comparisons of $\Delta$BIC for different models of time-resolved spectra for GRB 150510A.}
\begin{tabular}{cccccccc} 
\hline\hline
$t_{1}\sim t_{2}$&$\alpha$&$\quad$&$\quad$&$\Delta$BIC&$\quad$&$\quad$&\rm Best Model \\
(s)&$\quad$&Band&BB&PL+BB&CPL+BB&Band+BB& \\

\hline
0.00 	$\sim$	0.12 	&$	-0.68 	_{	-0.05 	}^{+	0.05 	}$&	19.06 	&	296.24 	&	-0.31 	&	Non-Con.	&	Non-Con.	&	PL+BB(CPL)	\\
0.12 	$\sim$	0.19 	&$	-0.57 	_{	-0.07 	}^{+	0.07 	}$&	3.90 	&	121.61 	&	-23.06 	&	Non-Con.	&	Non-Con.	&	PL+BB	\\
0.19 	$\sim$	0.28 	&$	-0.63 	_{	-0.06 	}^{+	0.06 	}$&	18.94 	&	253.02 	&	18.52 	&	Non-Con.	&	Non-Con.	&	CPL	\\
0.28 	$\sim$	0.48 	&$	-0.56 	_{	-0.06 	}^{+	0.06 	}$&	18.75 	&	245.32 	&	1.85 	&	Non-Con.	&	Non-Con.	&	CPL(PL+BB)	\\
0.48 	$\sim$	0.61 	&$	-0.50 	_{	-0.06 	}^{+	0.06 	}$&	23.21 	&	163.12 	&	6.65 	&	Non-Con.	&	Non-Con.	&	CPL	\\
0.61 	$\sim$	0.90 	&$	-0.62 	_{	-0.04 	}^{+	0.04 	}$&	28.00 	&	346.08 	&	34.72 	&	Non-Con.	&	Non-Con.	&	CPL	\\
\hline
\end{tabular}
\label{Table10}
\end{table*}


\begin{table*}
\centering
\renewcommand\tabcolsep{2.2pt}
\renewcommand\arraystretch{0.9}
\caption{Comparisons of $\Delta$BIC for different models of time-resolved spectra for GRB 170115B.}
\begin{tabular}{cccccccc} 
\hline\hline
$t_{1}\sim t_{2}$&$\alpha$&$\quad$&$\quad$&$\Delta$BIC&$\quad$&$\quad$&\rm Best Model \\
(s)&$\quad$&Band&BB&PL+BB&CPL+BB&Band+BB& \\

\hline
0.00 	$\sim$	0.54 	&$	-0.51 	_{	-0.04 	}^{+	0.04 	}$&	14.55 	&	257.24 	&	13.70 	&	Non-Con.	&	Non-Con.	&	CPL	\\
0.54 	$\sim$	1.22 	&$	-0.51 	_{	-0.04 	}^{+	0.04 	}$&	32.33 	&	302.73 	&	-5.95 	&	Non-Con.	&	Non-Con.	&	PL+BB(CPL)	\\
1.22 	$\sim$	1.54 	&$	-0.40 	_{	-0.05 	}^{+	0.05 	}$&	34.04 	&	206.20 	&	21.42 	&	Non-Con.	&	Non-Con.	&	CPL	\\
1.54 	$\sim$	2.23 	&$	-0.58 	_{	-0.04 	}^{+	0.04 	}$&	22.81 	&	226.55 	&	-0.07 	&	Non-Con.	&	Non-Con.	&	PL+BB(CPL)	\\
2.23 	$\sim$	5.08 	&$	-0.69 	_{	-0.05 	}^{+	0.05 	}$&	7.44 	&	242.03 	&	-39.99 	&	Non-Con.	&	Non-Con.	&	PL+BB	\\
5.08 	$\sim$	7.45 	&$	-0.68 	_{	-0.05 	}^{+	0.05 	}$&	13.63 	&	306.02 	&	-29.67 	&	Non-Con.	&	Non-Con.	&	PL+BB	\\
7.45 	$\sim$	8.44 	&$	-0.61 	_{	-0.05 	}^{+	0.05 	}$&	20.96 	&	208.46 	&	-11.36 	&	Non-Con.	&	Non-Con.	&	PL+BB	\\
8.44 	$\sim$	10.36 	&$	-0.69 	_{	-0.04 	}^{+	0.04 	}$&	21.78 	&	306.55 	&	-38.79 	&	Non-Con.	&	Non-Con.	&	PL+BB	\\
10.36 	$\sim$	11.42 	&$	-0.56 	_{	-0.05 	}^{+	0.05 	}$&	15.24 	&	202.39 	&	-19.85 	&	Non-Con.	&	Non-Con.	&	PL+BB	\\
11.42 	$\sim$	13.59 	&$	-0.73 	_{	-0.04 	}^{+	0.04 	}$&	13.39 	&	354.43 	&	-2.59 	&	Non-Con.	&	Non-Con.	&	PL+BB(CPL)	\\
13.59 	$\sim$	33.72 	&$	-1.07 	_{	-0.08 	}^{+	0.04 	}$&	Non-Con.	&	Non-Con.	&	-68.35 	&	Non-Con.	&	Non-Con.	&	PL+BB	\\

\hline
\end{tabular}
\label{Table11}
\end{table*}


\clearpage
\begin{thebibliography}{99}

\bibitem[Abdo et al.(2009)]{2009ApJ...706L.138A} Abdo, A.~A., Ackermann, M., Ajello, M., et al.\ 2009, \apjl, 706, L138. doi:10.1088/0004-637X/706/1/L138
\bibitem[Acuner et al.(2019)]{2019MNRAS.487.5508A} Acuner, Z., Ryde, F., \& Yu, H.-F.\ 2019, \mnras, 487, 5508. doi:10.1093/mnras/stz1356
\bibitem[Acuner et al.(2020)]{2020ApJ...893..128A} Acuner, Z., Ryde, F., Pe'er, A., et al.\ 2020, \apj, 893, 128. doi:10.3847/1538-4357/ab80c7
\bibitem[Arimoto et al.(2020)]{2020ApJ...891..106A} Arimoto, M., Asano, K., Tachibana, Y., et al.\ 2020, \apj, 891, 106. doi:10.3847/1538-4357/ab72f7
\bibitem[Asano \& M{\'e}sz{\'a}ros(2011)]{2011ApJ...739..103A} Asano, K. \& M{\'e}sz{\'a}ros, P.\ 2011, \apj, 739, 103. doi:10.1088/0004-637X/739/2/103
\bibitem[Asano \& Takahara(2007)]{2007ApJ...655..762A} Asano, K. \& Takahara, F.\ 2007, \apj, 655, 762. doi:10.1086/509756
\bibitem[Asano et al.(2009)]{2009ApJ...705L.191A} Asano, K., Guiriec, S., \& M{\'e}sz{\'a}ros, P.\ 2009, \apjl, 705, L191. doi:10.1088/0004-637X/705/2/L191
\bibitem[Axelsson et al.(2012)]{2012ApJ...757L..31A} Axelsson, M., Baldini, L., Barbiellini, G., et al.\ 2012, \apjl, 757, L31. doi:10.1088/2041-8205/757/2/L31
\bibitem[Band et al.(1993)]{1993ApJ...413..281B} Band, D., Matteson, J., Ford, L., et al.\ 1993, \apj, 413, 281. doi:10.1086/172995
\bibitem[Basak \& Rao(2014)]{2014MNRAS.442..419B} Basak, R. \& Rao, A.~R.\ 2014, \mnras, 442, 419. doi:10.1093/mnras/stu882
\bibitem[Beloborodov et al.(2014)]{2014ApJ...788...36B} Beloborodov, A.~M., Hasco{\"e}t, R., \& Vurm, I.\ 2014, \apj, 788, 36. doi:10.1088/0004-637X/788/1/36
\bibitem[Beloborodov(2010)]{2010MNRAS.407.1033B} Beloborodov, A.~M.\ 2010, \mnras, 407, 1033. doi:10.1111/j.1365-2966.2010.16770.x
\bibitem[Berger(2014)]{2014ARA&A..52...43B} Berger, E.\ 2014, \araa, 52, 43. doi:10.1146/annurev-astro-081913-035926
\bibitem[Blandford \& Znajek(1977)]{1977MNRAS.179..433B} Blandford, R.~D. \& Znajek, R.~L.\ 1977, \mnras, 179, 433. doi:10.1093/mnras/179.3.433
\bibitem[Bo{\v{s}}njak et al.(2009)]{2009A&A...498..677B} Bo{\v{s}}njak, {\v{Z}}., Daigne, F., \& Dubus, G.\ 2009, \aap, 498, 677. doi:10.1051/0004-6361/200811375
\bibitem[Borgonovo \& Ryde(2001)]{2001ApJ...548..770B} Borgonovo, L. \& Ryde, F.\ 2001, \apj, 548, 770. doi:10.1086/319008
\bibitem[Cash(1979)]{1979ApJ...228..939C} Cash, W.\ 1979, \apj, 228, 939. doi:10.1086/156922
\bibitem[Chang et al.(2023)]{2023ApJ...943..146C} Chang, X.-Z., L{\"u}, H.-J., Yang, X., et al.\ 2023, \apj, 943, 146. doi:10.3847/1538-4357/aca969
\bibitem[Chen et al.(2022)]{2022ApJ...932...25C} Chen, J.-M., Peng, Z.-Y., Du, T.-T., et al.\ 2022, \apj, 932, 25. doi:10.3847/1538-4357/ac6c2a
\bibitem[Corsi et al.(2010)]{2010A&A...524A..92C} Corsi, A., Guetta, D., \& Piro, L.\ 2010, \aap, 524, A92. doi:10.1051/0004-6361/200912461
\bibitem[Crider et al.(1997)]{1997ApJ...479L..39C} Crider, A., Liang, E.~P., Smith, I.~A., et al.\ 1997, \apjl, 479, L39. doi:10.1086/310574
\bibitem[Dai \& Lu(1998a)]{Dai1998a} Dai, Z.~G. \& Lu, T.\ 1998a, \mnras, 298, 87. doi:10.1046/j.1365-8711.1998.01681.x
\bibitem[Dai \& Lu(1998b)]{Dai1998b} Dai, Z.~G. \& Lu, T.\ 1998b, \aap, 333, L87
\bibitem[Daigne \& Mochkovitch(1998)]{1998MNRAS.296..275D} Daigne, F. \& Mochkovitch, R.\ 1998, \mnras, 296, 275. doi:10.1046/j.1365-8711.1998.01305.x
\bibitem[Della Valle et al.(2006)]{2006Natur.444.1050D} Della Valle, M., Chincarini, G., Panagia, N., et al.\ 2006, \nat, 444, 1050. doi:10.1038/nature05374
\bibitem[Du et al.(2024a)]{2024MNRAS.529L..67D} Du, Z.-W., L{\"u}, H., Liu, X., et al.\ 2024a, \mnras, 529, L67. doi:10.1093/mnrasl/slad203
\bibitem[Du et al.(2024b)]{2024ApJ...962L..27D} Du, Z., L{\"u}, H., Yuan, Y., et al.\ 2024b, \apjl, 962, L27. doi:10.3847/2041-8213/ad22e2
\bibitem[Feroz \& Skilling(2013)]{2013AIPC.1553..106F} Feroz, F. \& Skilling, J.\ 2013, Bayesian Inference and Maximum Entropy Methods in Science and Engineering: 32nd International Workshop on Bayesian Inference and Maximum Entropy Methods in Science and Engineering, 1553, 106. doi:10.1063/1.4819989
\bibitem[Feroz et al.(2009)]{2009MNRAS.398.1601F} Feroz, F., Hobson, M.~P., \& Bridges, M.\ 2009, \mnras, 398, 1601. doi:10.1111/j.1365-2966.2009.14548.x
\bibitem[Fraija et al.(2017)]{2017ApJ...848...94F} Fraija, N., Lee, W.~H., Araya, M., et al.\ 2017, \apj, 848, 94. doi:10.3847/1538-4357/aa8d65
\bibitem[Fruchter et al.(2006)]{2006Natur.441..463F} Fruchter, A.~S., Levan, A.~J., Strolger, L., et al.\ 2006, \nat, 441, 463. doi:10.1038/nature04787
\bibitem[Fynbo et al.(2006)]{2006Natur.444.1047F} Fynbo, J.~P.~U., Watson, D., Th{\"o}ne, C.~C., et al.\ 2006, \nat, 444, 1047. doi:10.1038/nature05375
\bibitem[Galama et al.(1998)]{1998Natur.395..670G} Galama, T.~J., Vreeswijk, P.~M., van Paradijs, J., et al.\ 1998, \nat, 395, 670. doi:10.1038/27150
\bibitem[Gal-Yam et al.(2006)]{2006Natur.444.1053G} Gal-Yam, A., Fox, D.~B., Price, P.~A., et al.\ 2006, \nat, 444, 1053. doi:10.1038/nature05373
\bibitem[Gao \& Zhang(2015)]{2015ApJ...801...2} Gao, H. \& Zhang B et al.\ 2015, \apj, 801, 5. doi:0.1088/0004-637X/801/2/103
\bibitem[Gehrels et al.(2006)]{2006Natur.444.1044G} Gehrels, N., Norris, J.~P., Barthelmy, S.~D., et al.\ 2006, \nat, 444, 1044. doi:10.1038/nature05376
\bibitem[Golenetskii et al.(1983)]{1983Natur.306..451G} Golenetskii, S.~V., Mazets, E.~P., Aptekar, R.~L., et al.\ 1983, \nat, 306, 451. doi:10.1038/306451a0
\bibitem[Gompertz et al.(2023)]{2023NatAs...7...67G} Gompertz, B.~P., Ravasio, M.~E., Nicholl, M., et al.\ 2023, Nature Astronomy, 7, 67. doi:10.1038/s41550-022-01819-4
\bibitem[Goodman \& Weare(2010)]{2010CAMCS...5...65G} Goodman, J. \& Weare, J.\ 2010, Communications in Applied Mathematics and Computational Science, 5, 65. doi:10.2140/camcos.2010.5.65
\bibitem[Goodman(1986)]{1986ApJ...308L..47G} Goodman, J.\ 1986, \apjl, 308, L47. doi:10.1086/184741
\bibitem[Gu et al.(2006)]{2006ApJ...643L..87G} Gu, W.-M., Liu, T., \& Lu, J.-F.\ 2006, \apjl, 643, L87. doi:10.1086/505140
\bibitem[Guiriec et al.(2011)]{2011ApJ...727L..33G} Guiriec, S., Connaughton, V., Briggs, M.~S., et al.\ 2011, \apjl, 727, L33. doi:10.1088/2041-8205/727/2/L33
\bibitem[Guiriec et al.(2013)]{2013ApJ...770...32G} Guiriec, S., Daigne, F., Hasco{\"e}t, R., et al.\ 2013, \apj, 770, 32. doi:10.1088/0004-637X/770/1/32
\bibitem[Guiriec et al.(2015)]{2015ApJ...807..148G} Guiriec, S., Kouveliotou, C., Daigne, F., et al.\ 2015, \apj, 807, 148. doi:10.1088/0004-637X/807/2/148
\bibitem[Gupta \& Zhang(2007)]{2007MNRAS.380...78G} Gupta, N. \& Zhang, B.\ 2007, \mnras, 380, 78. doi:10.1111/j.1365-2966.2007.12051.x
\bibitem[Hou et al.(2018)]{2018ApJ...866...13H} Hou, S.-J., Zhang, B.-B., Meng, Y.-Z., et al.\ 2018, \apj, 866, 13. doi:10.3847/1538-4357/aadc07
\bibitem[Ioka(2010a)]{2010AIPC.1279...28I} Ioka, K.\ 2010a, Deciphering the Ancient Universe with Gamma-ray Bursts, 1279, 28. doi:10.1063/1.3509286
\bibitem[Ioka(2010b)]{2010PThPh.124..667I} Ioka, K.\ 2010b, Progress of Theoretical Physics, 124, 667. doi:10.1143/PTP.124.667
\bibitem[Kaneko et al.(2015)]{2015MNRAS.452..824K} Kaneko, Y., Bostanc{\i}, Z.~F., G{\"o}{\u{g}}{\"u}{\c{s}}, E., et al.\ 2015, \mnras, 452, 824. doi:10.1093/mnras/stv1286
\bibitem[Kargatis et al.(1994)]{1994ApJ...422..260K} Kargatis, V.~E., Liang, E.~P., Hurley, K.~C., et al.\ 1994, \apj, 422, 260. doi:10.1086/173724
\bibitem[Kobayashi et al.(1997)]{1997ApJ...490...92K} Kobayashi, S., Piran, T., \& Sari, R.\ 1997, \apj, 490, 92. doi:10.1086/512791
\bibitem[Kouveliotou et al.(1993)]{1993ApJ...413L.101K} Kouveliotou, C., Meegan, C.~A., Fishman, G.~J., et al.\ 1993, \apjl, 413, L101. doi:10.1086/186969
\bibitem[Kumar \& Barniol Duran(2009)]{2009MNRAS.400L..75K} Kumar, P. \& Barniol Duran, R.\ 2009, \mnras, 400, L75. doi:10.1111/j.1745-3933.2009.00766.x
\bibitem[Kumar \& Zhang(2015)]{2015PhR...561....1K} Kumar, P. \& Zhang, B.\ 2015, \physrep, 561, 1. doi:10.1016/j.physrep.2014.09.008
\bibitem[L{\"u} \& Zhang(2014)]{Lv2014} L{\"u}, H.-J. \& Zhang, B.\ 2014, \apj, 785, 74. doi:10.1088/0004-637X/785/1/74
\bibitem[L{\"u} et al.(2014)]{2014MNRAS.442.1922L} L{\"u}, H.-J., Zhang, B., Liang, E.-W., et al.\ 2014, \mnras, 442, 1922. doi:10.1093/mnras/stu982
\bibitem[L{\"u} et al.(2015)]{Lv2015} L{\"u}, H.-J., Zhang, B., Lei, W.-H., et al.\ 2015, \apj, 805, 89. doi:10.1088/0004-637X/805/2/89
\bibitem[L{\"u} et al.(2017)]{2017ApJ...849...71L} L{\"u}, H.-J., L{\"u}, J., Zhong, S.-Q., et al.\ 2017, \apj, 849, 71. doi:10.3847/1538-4357/aa8f99
\bibitem[L{\"u} et al.(2022)]{2022ApJ...931L..23L} L{\"u}, H.-J., Yuan, H.-Y., Yi, T.-F., et al.\ 2022, \apjl, 931, L23. doi:10.3847/2041-8213/ac6e3a
\bibitem[Lan et al.(2018)]{2018ApJ...862..155L} Lan, L., L{\"u}, H.-J., Zhong, S.-Q., et al.\ 2018, \apj, 862, 155. doi:10.3847/1538-4357/aacda6
\bibitem[Lan et al.(2020)]{2020MNRAS.492.3622L} Lan, L., Lu, R.-J., L{\"u}, H.-J., et al.\ 2020, \mnras, 492, 3622. doi:10.1093/mnras/staa044
\bibitem[Lazzati \& Begelman(2010)]{2010ApJ...725.1137L} Lazzati, D. \& Begelman, M.~C.\ 2010, \apj, 725, 1137. doi:10.1088/0004-637X/725/1/1137
\bibitem[Lei et al.(2013)]{2013ApJ...765..125L} Lei, W.-H., Zhang, B., \& Liang, E.-W.\ 2013, \apj, 765, 125. doi:10.1088/0004-637X/765/2/125
\bibitem[Lei et al.(2017)]{2017ApJ...849...47L} Lei, W.-H., Zhang, B., Wu, X.-F., et al.\ 2017, \apj, 849, 47. doi:10.3847/1538-4357/aa9074
\bibitem[Li(2019)]{2019ApJS..242...16L} Li, L.\ 2019, \apjs, 242, 16. doi:10.3847/1538-4365/ab1b78
\bibitem[Li(2020)]{2020ApJ...894..100L} Li, L.\ 2020, \apj, 894, 100. doi:10.3847/1538-4357/ab8014
\bibitem[Li et al.(2023)]{2023ApJ...944L..57L} Li, L., Wang, Y., Ryde, F., et al.\ 2023, \apjl, 944, L57. doi:10.3847/2041-8213/acb99d
\bibitem[Lloyd-Ronning \& Petrosian(2002)]{2002ApJ...565..182L} Lloyd-Ronning, N.~M. \& Petrosian, V.\ 2002, \apj, 565, 182. doi:10.1086/324484
\bibitem[Lu et al.(2012)]{2012ApJ...756..112L} Lu, R.-J., Wei, J.-J., Liang, E.-W., et al.\ 2012, \apj, 756, 112. doi:10.1088/0004-637X/756/2/112
\bibitem[Medvedev(2000)]{2000ApJ...540..704M} Medvedev, M.~V.\ 2000, \apj, 540, 704. doi:10.1086/309374
\bibitem[M{\'e}sz{\'a}ros et al.(1993)]{1993ApJ...415..181M}M{\'e}sz{\'a}ros, P., Laguna, P., \& Rees, M.~J.\ 1993, \apj, 415, 181. doi:10.1086/173154
\bibitem[M{\'e}sz{\'a}ros \& Rees(1994)]{1994MNRAS.269L..41M} M{\'e}sz{\'a}ros, P. \& Rees, M.~J.\ 1994, \mnras, 269, L41. doi:10.1093/mnras/269.1.L41
\bibitem[M{\'e}sz{\'a}ros \& Rees(1997)]{1997ApJ...476..232M} M{\'e}sz{\'a}ros, P. \& Rees, M.~J.\ 1997, \apj, 476, 232. doi:10.1086/303625
\bibitem[M{\'e}sz{\'a}ros \& Rees(2000)]{2000ApJ...530..292M} M{\'e}sz{\'a}ros, P. \& Rees, M.~J.\ 2000, \apj, 530, 292. doi:10.1086/308371
\bibitem[Meegan et al.(2009)]{2009ApJ...702..791M} Meegan, C., Lichti, G., Bhat, P.~N., et al.\ 2009, \apj, 702, 791. doi:10.1088/0004-637X/702/1/791
\bibitem[Narayan et al.(2001)]{Narayan2001} Narayan, R., Piran, T., \& Kumar, P.\ 2001, \apj, 557, 949. doi:10.1086/322267
\bibitem[Neath \& Cavanaugh(2012)]{Neath2012} Neath, A.~A. \& Cavanaugh, J.~E.\ 2012, WIREs Comput. Stat., 4, 199. doi: 10.1002/wics.199
\bibitem[Norris et al.(2010)]{2010ApJ...717..411N} Norris, J.~P., Gehrels, N., \& Scargle, J.~D.\ 2010, \apj, 717, 411. doi:10.1088/0004-637X/717/1/411
\bibitem[Paczynski(1986)]{1986ApJ...308L..43P} Paczynski, B.\ 1986, \apjl, 308, L43. doi:10.1086/184740
\bibitem[Panaitescu \& M{\'e}sz{\'a}ros(2000)]{2000ApJ...544L..17P} Panaitescu, A. \& M{\'e}sz{\'a}ros, P.\ 2000, \apjl, 544, L17. doi:10.1086/317301
\bibitem[Pe'er \& Ryde(2011)]{2011ApJ...732...49P} Pe'er, A. \& Ryde, F.\ 2011, \apj, 732, 49. doi:10.1088/0004-637X/732/1/49
\bibitem[Pe'er et al.(2006)]{2006AIPC..836..181P} Pe'er, A., M{\'e}sz{\'a}ros, P., \& Rees, M.~J.\ 2006, Gamma-Ray Bursts in the Swift Era, 836, 181. doi:10.1063/1.2207885
\bibitem[Pe'er et al.(2012)]{2012MNRAS.420..468P} Pe'er, A., Zhang, B.-B., Ryde, F., et al.\ 2012, \mnras, 420, 468. doi:10.1111/j.1365-2966.2011.20052.x
\bibitem[Peng et al.(2024)]{2024ApJ...969...26P} Peng, Z.-Y., Chen, J.-M., \& Mao, J.\ 2024, \apj, 969, 26. doi:10.3847/1538-4357/ad45fc
\bibitem[Piran et al.(1993)]{1993MNRAS.263..861P} Piran, T., Shemi, A., \& Narayan, R.\ 1993, \mnras, 263, 861. doi:10.1093/mnras/263.4.861
\bibitem[Popham et al.(1999)]{1999ApJ...518..356P} Popham, R., Woosley, S.~E., \& Fryer, C.\ 1999, \apj, 518, 356. doi:10.1086/307259
\bibitem[Preece et al.(1998)]{1998ApJ...506L..23P} Preece, R.~D., Briggs, M.~S., Mallozzi, R.~S., et al.\ 1998, \apjl, 506, L23. doi:10.1086/311644
\bibitem[Rastinejad et al.(2022)]{2022Natur.612..223R} Rastinejad, J.~C., Gompertz, B.~P., Levan, A.~J., et al.\ 2022, \nat, 612, 223. doi:10.1038/s41586-022-05390-w
\bibitem[Rees \& M{\'e}sz{\'a}ros(1994)]{1994ApJ...430L..93R} Rees, M.~J. \& Meszaros, P.\ 1994, \apjl, 430, L93. doi:10.1086/187446
\bibitem[Rees \& M{\'e}sz{\'a}ros(2005)]{2005ApJ...628..847R} Rees, M.~J. \& M{\'e}sz{\'a}ros, P.\ 2005, \apj, 628, 847. doi:10.1086/430818f
\bibitem[Ryde et al.(2010)]{2010ApJ...709L.172R} Ryde, F., Axelsson, M., Zhang, B.~B., et al.\  2010, \apjl, 709, L172. doi:10.1088/2041-8205/709/2/L172
\bibitem[Ryde et al.(2019)]{2019MNRAS.484.1912R} Ryde, F., Yu, H.-F., Dereli-B{\'e}gu{\'e}, H., et al.\ 2019, \mnras, 484, 1912. doi:10.1093/mnras/stz083
\bibitem[Sakamoto et al.(2011)]{2011ApJS..195....2S} Sakamoto, T., Barthelmy, S.~D., Baumgartner, W.~H., et al.\ 2011, \apjs, 195, 2. doi:10.1088/0067-0049/195/1/2
\bibitem[Sari et al.(1998)]{1998ApJ...497L..17S} Sari, R., Piran, T., \& Narayan, R.\ 1998, \apjl, 497, L17. doi:10.1086/311269
\bibitem[Shemi \& Piran(1990)]{1990ApJ...365L..55S} Shemi, A. \& Piran, T.\ 1990, \apjl, 365, L55. doi:10.1086/185887
\bibitem[Stanek et al.(2003)]{2003ApJ...591L..17S} Stanek, K.~Z., Matheson, T., Garnavich, P.~M., et al.\ 2003, \apjl, 591, L17. doi:10.1086/376976
\bibitem[Sun et al.(2023)]{2023arXiv230705689S} Sun, H., Wang, C.-W., Yang, J., et al.\ 2023, arXiv:2307.05689. doi:10.48550/arXiv.2307.05689
\bibitem[Tang et al.(2017)]{2017ApJ...844...56T} Tang, Q.-W., Wang, X.-Y., \& Liu, R.-Y.\ 2017, \apj, 844, 56. doi:10.3847/1538-4357/aa7a58
\bibitem[Tang et al.(2021)]{2021ApJ...922..255T} Tang, Q.-W., Wang, K., Li, L., et al.\ 2021, \apj, 922, 255. doi:10.3847/1538-4357/ac26ba
\bibitem[Thompson(1994)]{Thompson1994} Thompson, C.\ 1994, \mnras, 270, 480. doi:10.1093/mnras/270.3.480
\bibitem[Troja et al.(2017)]{2017Natur.547..425T} Troja, E., Lipunov, V.~M., Mundell, C.~G., et al.\ 2017, \nat, 547, 425. doi:10.1038/nature23289
\bibitem[Troja et al.(2022)]{2022Natur.612..228T} Troja, E., Fryer, C.~L., O'Connor, B., et al.\ 2022, \nat, 612, 228. doi:10.1038/s41586-022-05327-3
\bibitem[Usov(1992)]{Usov1992} Usov, V.~V.\ 1992, \nat, 357, 472. doi:10.1038/357472a0
\bibitem[Vianello et al.(2017)]{2017ifs..confE.130V} Vianello, G., Lauer, R.~J., Burgess, J.~M., et al.\ 2017, Proceedings of the 7th International Fermi Symposium, 130
\bibitem[Vianello(2018)]{2018ApJS..236...17V} Vianello, G.\ 2018, \apjs, 236, 17. doi:10.3847/1538-4365/aab780
\bibitem[von Kienlin et al.(2020)]{2020ApJ...893...46V} von Kienlin, A., Meegan, C.~A., Paciesas, W.~S., et al.\ 2020, \apj, 893, 46. doi:10.3847/1538-4357/ab7a18
\bibitem[Wang et al.(2018)]{2018ApJ...857...24W} Wang, K., Liu, R.-Y., Dai, Z.-G., et al.\ 2018, \apj, 857, 24. doi:10.3847/1538-4357/aab667
\bibitem[Yang et al.(2015)]{2015NatCo...6.7323Y} Yang, B., Jin, Z.-P., Li, X., et al.\ 2015, Nature Communications, 6, 7323. doi:10.1038/ncomms8323
\bibitem[Yang et al.(2022)]{2022Natur.612..232Y} Yang, J., Ai, S., Zhang, B.-B., et al.\ 2022, \nat, 612, 232. doi:10.1038/s41586-022-05403-8
\bibitem[Yi et al.(2023)]{2023arXiv231007205Y} Yi, S.-X., Wang, C.-W., Shao, X.-Y., et al.\ 2023, arXiv:2310.07205. doi:10.48550/arXiv.2310.07205
\bibitem[Yu et al.(2016)]{2016A&A...588A.135Y} Yu, H.-F., Preece, R.~D., Greiner, J., et al.\ 2016, \aap, 588, A135. doi:10.1051/0004-6361/201527509
\bibitem[Zhang \& M{\'e}sz{\'a}ros(2001)]{Zhang2001} Zhang, B. \& M{\'e}sz{\'a}ros, P.\ 2001, \apjl, 552, L35. doi:10.1086/320255
\bibitem[Zhang \& Pe'er(2009)]{2009ApJ...700L..65Z} Zhang, B. \& Pe'er, A.\ 2009, \apjl, 700, L65. doi:10.1088/0004-637X/700/2/L65
\bibitem[Zhang \& Yan(2011)]{2011ApJ...726...90Z} Zhang, B. \& Yan, H.\ 2011, \apj, 726, 90. doi:10.1088/0004-637X/726/2/90
\bibitem[Zhang et al.(2009)]{2009ApJ...703.1696Z} Zhang, B., Zhang, B.-B., Virgili, F.~J., et al.\ 2009, \apj, 703, 1696. doi:10.1088/0004-637X/703/2/1696
\bibitem[Zhang(2011)]{2011CRPhy..12..206Z} Zhang, B.\ 2011, Comptes Rendus Physique, 12, 206. doi:10.1016/j.crhy.2011.03.004
\bibitem[Zhang(2014)]{2014AcASn..55..354Z} Zhang, B.\ 2014, Acta Astronomica Sinica, 55, 354
\bibitem[Zhang(2018)]{2018pgrb.book.....Z} Zhang, B.\ 2018, The Physics of Gamma-Ray Bursts by Bing Zhang. ISBN: 978-1-139-22653-0. Cambridge University Press, 2018. doi:10.1017/9781139226530
\bibitem[Zhang et al.(2011)]{2011ApJ...730..141Z} Zhang, B.-B., Zhang, B., Liang, E.-W., et al.\ 2011, \apj, 730, 141. doi:10.1088/0004-637X/730/2/141
\bibitem[Zhang et al.(2018)]{2018NatAs...2...69Z} Zhang, B.-B., Zhang, B., Castro-Tirado, A.~J., et al.\ 2018, Nature Astronomy, 2, 69. doi:10.1038/s41550-017-0309-8
\bibitem[Zhong et al.(2023)]{2023ApJ...947L..21Z} Zhong, S.-Q., Li, L., \& Dai, Z.-G.\ 2023, \apjl, 947, L21. doi:10.3847/2041-8213/acca83

\end{thebibliography}
\end{document}